\documentclass[twocolumn,trackchanges]{aastex701}

\shorttitle{Black Hole Scaling Relations}
\shortauthors{Contini et al.}

\begin{document}

\title{On The Scaling Relations Between Black Holes And The Galaxy/Bulge Mass And Their Connections With The Black Hole Occupation Fraction}

\author{Emanuele Contini}
\altaffiliation{Yonsei University}
\affiliation{Department of Astronomy and Yonsei University Observatory, Yonsei University, 50 Yonsei-ro, Seodaemun-gu, Seoul 03722, Republic of Korea}
\email[show]{emanuele.contini82@google.com}  

\author{J.K. Jang}
\altaffiliation{Yonsei University}
\affiliation{Department of Astronomy and Yonsei University Observatory, Yonsei University, 50 Yonsei-ro, Seodaemun-gu, Seoul 03722, Republic of Korea}
\email{starbrown816@gmail.com}
\author{Jinsu Rhee}
\altaffiliation{Institut d’Astrophysique de Paris}
\affiliation{Institut d’Astrophysique de Paris, Sorbonne Université, CNRS, UMR 7095, 98 bis bd Arago, 75014 Paris, France}
\email{emanuele.contini82@google.com}
\author{Changjo Seo}
\altaffiliation{Yonsei University}
\affiliation{Department of Astronomy and Yonsei University Observatory, Yonsei University, 50 Yonsei-ro, Seodaemun-gu, Seoul 03722, Republic of Korea}
\email{emanuele.contini82@gmail.com}
\author{Sukyoung K. Yi}
\altaffiliation{Yonsei University}
\affiliation{Department of Astronomy and Yonsei University Observatory, Yonsei University, 50 Yonsei-ro, Seodaemun-gu, Seoul 03722, Republic of Korea}
\email[show]{yi@yonsei.ac.kr}

\begin{abstract}
We investigate the black hole (BH)--host galaxy scaling relations predicted by the semianalytic model \texttt{FEGA25}, focusing on their dependence on BH mass threshold, galaxy morphology, and redshift, and on their connection with the BH occupation fraction (BHOF). Using the YS200 cosmological simulation as our reference volume, we study the $M_{\rm BH}$--$M_{\star}$ and $M_{\rm BH}$--$M_{\rm bulge}$ relations for different BH mass selections and compare their redshift evolution with the hydrodynamical zoom-in simulation \texttt{NewCluster}. We find that BH mass cuts primarily affect the low-mass regime, while the selected relations progressively converge toward the intrinsic no-cut relations at higher host masses. The $M_{\rm BH}$--$M_{\star}$ relation also depends systematically on galaxy morphology, with increasingly bulge-dominated systems preferentially occupying higher stellar masses. Over $z=0$--$3$, FEGA25 shows no clear systematic evolution of either scaling relation, consistent with a broadly preserved statistical connection between BH and host-galaxy growth. NewCluster shows no clear monotonic evolution either, although it exhibits larger scatter, particularly toward high host masses. We further show that BHOF and BH--host scaling relations encode complementary information. A low BHOF generally corresponds to a stronger selection-induced offset, but the mapping is not unique: substantially different occupation fractions can yield similar median scaling relations, and at fixed BHOF the bias depends on the host property and underlying BH mass distribution. At $z=0$, \texttt{FEGA25} is consistent with available observational constraints for both stellar- and bulge-mass relations, while predicting increased scatter and bending toward lower host masses.
\end{abstract}



\section{Introduction}
\label{sec:intro}

The existence of correlations between the mass of central black holes (BHs) and the properties of their host galaxies represents one of the main observational constraints on the joint evolution of galaxies and their nuclear BHs. Early studies established correlations between BH mass and bulge properties \citep{magorrian1998,marconi2003}, stellar velocity dispersion \citep{ferrarese2000,gebhardt2000,tremaine2002,gultekin2009}, and spheroid luminosity or stellar mass \citep{haring2004,sani2011,mcconnell2013,kormendy2013,lasker2014}. These relations have since become fundamental benchmarks for models of galaxy formation and BH growth and are commonly used to connect BH demographics with the assembly histories of their host galaxies.

Their interpretation, however, is not straightforward. Relations based on total stellar mass and bulge stellar mass do not necessarily trace the same physical connection, particularly in late-type and low-mass systems where the bulge may represent only a small fraction of the total stellar component. Observational samples show substantial differences in slope, normalization, and scatter depending on galaxy population, structural decomposition, and sample selection \citep{reines2015,bentz2018,sturm2024,sani2011,lasker2014,savorgnan2016}. In particular, AGN hosts can lie systematically below scaling relations defined by massive quiescent galaxies with dynamically measured BHs \citep{reines2015,sturm2024}. Galaxy morphology further affects these relations, with different structural populations occupying distinct regions of the BH--host scaling planes \citep{kormendy2011,kormendy2013,mcconnell2013,graham2013,davis2018,davis2019,sahu2019}. Studies separating S\'ersic and core-S\'ersic systems, or early- and late-type galaxies, have indeed found substantially different slopes and normalizations \citep{graham2013,scott2013,savorgnan2016,sahu2019}. In our previous implementation of the \texttt{FEGA} framework, where positive AGN feedback was introduced, \citet{contini2024d} similarly showed that the predicted $M_{\rm BH}$--$M_{\star}$ relation depends on morphology and, in particular, on the bulge-to-total stellar mass ratio.

Selection effects provide an additional complication. Samples of dynamically measured BHs may preferentially include galaxies with high velocity dispersions and massive BHs, potentially biasing both the normalization and slope of the inferred relations \citep{gultekin2009,lauer2007,shankar2016,shankar2017,shankar2019}. AGN-selected samples are affected by a different combination of luminosity limits, duty cycles, Eddington-ratio distributions, obscuration, intrinsic scatter, and the requirement that the BH be actively accreting \citep{lauer2007,schulze2011,schulze2014}. Numerical experiments have likewise shown that selections based on BH or host properties can produce scaling relations that differ substantially from those of the intrinsic population \citep{degraf2015}. Consequently, apparently different BH--host relations need not reflect differences in BH-growth physics alone, and comparisons between samples require their underlying selection functions to be considered explicitly.

The redshift evolution of these relations remains similarly uncertain. Studies extending from the local Universe to $z\sim1$--$2$ have often suggested that BHs may be more massive at fixed stellar or spheroid mass at earlier epochs \citep{peng2006,merloni2010,bennert2011,ding2020}, while other analyses find much weaker evolution, particularly when total rather than bulge stellar mass is considered \citep{jahnke2009,suh2020}. Moreover, accounting explicitly for selection effects can substantially reduce the statistical evidence for evolution \citep{schulze2011,schulze2014}. JWST has now extended these measurements toward lower host masses and much earlier cosmic epochs, while also enabling direct decomposition of AGN and host emission at redshifts that were extremely challenging with previous facilities \citep{ding2022}. Several high-redshift ($z>3$) AGN samples appear to contain BHs that are overmassive relative to local relations \citep{pacucci2023,mezcua2024}, whereas other analyses show that observational uncertainties and selection effects can substantially reduce the apparent discrepancy \citep{tanaka2024,li2024}. This problem is closely connected to the detection of rapidly growing BHs at very early times \citep{larson2023,maiolino2024,natarajan2024,banados2018,wang2021}, which places strong constraints on BH formation and subsequent growth.

Theoretical models predict broadly similar BH--host correlations but differ significantly in their detailed properties. Early cosmological hydrodynamical calculations already showed that BH growth and feedback can generate correlations between BH and galaxy properties resembling the observed local relations \citep{dimatteo2008}. More recent large-volume simulations, including MassiveBlack-II \citep{khandai2015}, IllustrisTNG \citep{pillepich2018}, EAGLE \citep{schaye2015}, Horizon-AGN \citep{dubois2014}, SIMBA \citep{dave2019} and COLIBRE \citep{schaye2026}, predict substantial variations in the normalization, slope, scatter, and redshift evolution of BH--host relations as a consequence of different BH seeding, accretion, dynamics, and feedback implementations \citep{khandai2015,sijacki2015,schaye2015,rosasguevara2016,dubois2016,weinberger2018,dave2019,thomas2019,li2020,habouzit2021,terrazas2020}. The predicted relations can also depend on galaxy morphology and star-formation state \citep{sijacki2015,dubois2016}, further emphasizing that the global $M_{\rm BH}$--$M_{\star}$ relation represents the combination of physically distinct galaxy populations. These differences become particularly important toward low galaxy masses, where BH populations retain a stronger sensitivity to numerical resolution, BH dynamics, and the adopted seeding model \citep{bellovary2019,haidar2022,bhowmick2025,bhowmick2026}.

Semianalytic models (SAMs) provide a complementary approach and have been extensively used to investigate BH growth, AGN feedback, seed formation, and the resulting BH--galaxy connection \citep{somerville2008,lamastra2010,barausse2012,dayal2018,fontanot2020,trinca2022,spinoso2023,cammelli2025,contini2024d,contini2025,contini2025AGN,bonoli2025}. Their ability to follow large galaxy populations over extended ranges in mass and redshift makes them particularly useful for separating the effects of BH-growth prescriptions from those associated with galaxy assembly and sample demographics. Within this framework, \texttt{FEGA25} \citep{contini2025,contini2025AGN} follows the coupled evolution of galaxies and BHs while accounting for multiple AGN-feedback channels.

A complementary description of the BH population is provided by the BH occupation fraction (BHOF), $f_{\rm BH,occ}$, defined as the fraction of galaxies hosting a BH above a given mass threshold. The BHOF is particularly informative at low stellar masses, where galaxies and their BHs may retain information about BH formation and early growth \citep{volonteri-natarajan2009,greene2012,miller2015,bellovary2019,tremmel2024,bhowmick2025}. Theoretical predictions span a broad range because of differences in BH seeding, accretion, dynamics, feedback, environment, and numerical resolution \citep{haidar2022,tremmel2024,bhowmick2025,bhowmick2026}. SAMs similarly predict a dependence on the adopted BH formation and growth channels \citep{dayal2018,trinca2022,spinoso2023,bonoli2025}. Observational estimates remain uncertain, particularly in dwarf galaxies. Searches based on optical and X-ray signatures have revealed sizable populations of accreting BH candidates in low-mass hosts \citep{gallo2010,reines2013,mezcua2018}, but translating active fractions into an intrinsic BHOF requires assumptions about accretion rates and detectability. Current studies therefore continue to yield substantially different estimates of the local BHOF \citep{miller2015,burke2025,zou2025}. The comparison is further complicated by the fact that observational studies generally identify actively accreting BHs rather than measuring the intrinsic occupation fraction directly.

In our previous work (\citealt{contini2026BH}, hereafter C26), we used \texttt{FEGA25} to show that the BHOF--stellar mass relation depends not only on stellar mass, but also on BH mass threshold, galaxy type, numerical resolution, simulated volume, sampled galaxy population, and redshift. C26 therefore demonstrated that $f_{\rm BH,occ}$ cannot generally be interpreted as a universal function of stellar mass alone.

Here we address the complementary question of how this demographic incompleteness affects the BH--host scaling relations themselves. We investigate the $M_{\rm BH}$--$M_{\star}$ and $M_{\rm BH}$--$M_{\rm bulge}$ relations predicted by \texttt{FEGA25}, their dependence on BH mass threshold, morphology, and redshift, and their connection with the corresponding BHOF. We further quantify the bias induced by BH-mass selection, compare the local predictions with observational constraints, and use the hydrodynamical zoom-in simulation \texttt{NewCluster} \citep{han2026} as an independent comparison of the redshift evolution. Our aim is therefore to connect two complementary descriptions of BH demographics: the properties of the selected BH population and the fraction of galaxies that this population represents.

\section{Methods}
\label{sec:methods}

\subsection{\texttt{FEGA25} and YS200}

\texttt{FEGA25} is the latest implementation of the \textit{Formation and Evolution of GAlaxies} semianalytic framework \citep{contini2025,contini2025AGN}. It is an updated version of the model presented in \citet{contini2024d} and includes a revised treatment of star formation and AGN feedback. In particular, the model adopts an extended Kennicutt--Schmidt relation \citep{shi2011} and a self-consistent coupling between BH growth and AGN activity. BHs grow through merger-driven cold-gas accretion and BH coalescence, following the classical quasar-mode prescription of \citet{croton2006}, together with subsequent radio-mode accretion from the hot gas reservoir. We refer the reader to C26 and the references therein for a detailed description of these prescriptions and of the calibration of \texttt{FEGA25}.

While C26 investigated the dependence of the BH occupation fraction on simulation volume, numerical resolution, and sampled galaxy population using three cosmological boxes, here we restrict the semianalytic analysis to YS200 (in order to avoid limitations due to resolution). The dependence on these effects was already explored in C26, allowing us to focus here on the BH--host galaxy scaling relations using a single reference cosmological volume.

YS200 is a dark-matter-only cosmological simulation with a box size of $200\,{\rm Mpc}\,h^{-1}$ and a dark matter particle mass of $3.26\times10^{8}\,M_\odot h^{-1}$. The simulation was performed with \textsc{Gadget-4} \citep{springel2021}, assuming the Planck 2018 cosmology \citep{planck2020}, and its halo merger trees are populated with galaxies using \texttt{FEGA25}. Stellar masses are derived assuming a Chabrier initial mass function \citep{chabrier2003}.

As in C26, we adopt YS200 galaxy catalog, requiring $\log(M_{\star}/M_\odot)>7$ and a minimum host halo mass of $\log(M_{\rm halo}/M_\odot)=12.51$. This corresponds approximately to halos resolved with at least $10^{4}$ dark matter particles and is designed to minimize numerical effects associated with poorly resolved systems.

\subsection{\texttt{NewCluster}}

We complement the semianalytic predictions with \texttt{NewCluster} \citep{han2026}, a cosmological hydrodynamical zoom-in simulation designed to study galaxy formation within a massive cluster environment. The simulation follows the evolution of a galaxy cluster with virial mass $\sim5\times10^{14}\,M_\odot$ at $z=0$.

The simulation was performed with the \textsc{RAMSES-yOMP} code \citep{han2025a}, a hybrid-parallel implementation of the adaptive-mesh-refinement hydrodynamics solver \textsc{RAMSES} \citep{teyssier2002}. The parent simulation box has a size of $100\,{\rm Mpc}\,h^{-1}$, while the zoomed region, at $z=0$, extends to approximately $3.5R_{\rm vir}$ around the central cluster. Within the high-resolution region, the dark matter particle mass is $m_{\rm DM}=1.3\times10^{6}\,M_\odot$, while stellar particles have a mass of $m_\star \sim2\times10^{4}\,M_\odot$. The best spatial resolution is approximately $70$ pc in comoving units.

The simulation includes radiative cooling, star formation, chemical enrichment, SN feedback, and AGN feedback. The latter follows the dual-mode scheme of \citet{dubois2012}, with kinetic jet feedback operating at low Eddington ratios and thermal quasar-mode feedback active at high accretion rates. \texttt{NewCluster} adopts a BH seed mass of $10^{5}\,M_\odot$; we refer to \citet{han2026} and C26 for a complete description of the simulation and its BH implementation.

For the present analysis, the bulge component in \texttt{NewCluster} is identified from the radial stellar surface-density profile measured from the face-on projection of each galaxy. The profile, constructed from the centre out to twice the half mass radius $R_{50}$, is decomposed into a S\'ersic bulge and an exponential disk, with the disk S\'ersic index fixed to $n=1$ and the bulge index allowed to vary between $0.5$ and $10$. When the two fitted components intersect, their intersection defines $r_{\rm cross}$, and the bulge properties are measured within this radius.

For the present analysis, we use the total stellar mass, bulge stellar mass, and mass of the most massive BH associated with each galaxy provided in the \texttt{NewCluster} catalogs. These quantities are used to construct the $M_{\rm BH}$--$M_{\star}$ and $M_{\rm BH}$--$M_{\rm bulge}$ relations shown below.

\section{Results And Discussion}
\label{sec:results}

\subsection{Stellar/Bulge vs Black Hole Mass Relations}
\label{sec:relations}

\begin{figure*}[t!]
\centering
\includegraphics[width=0.94\textwidth]{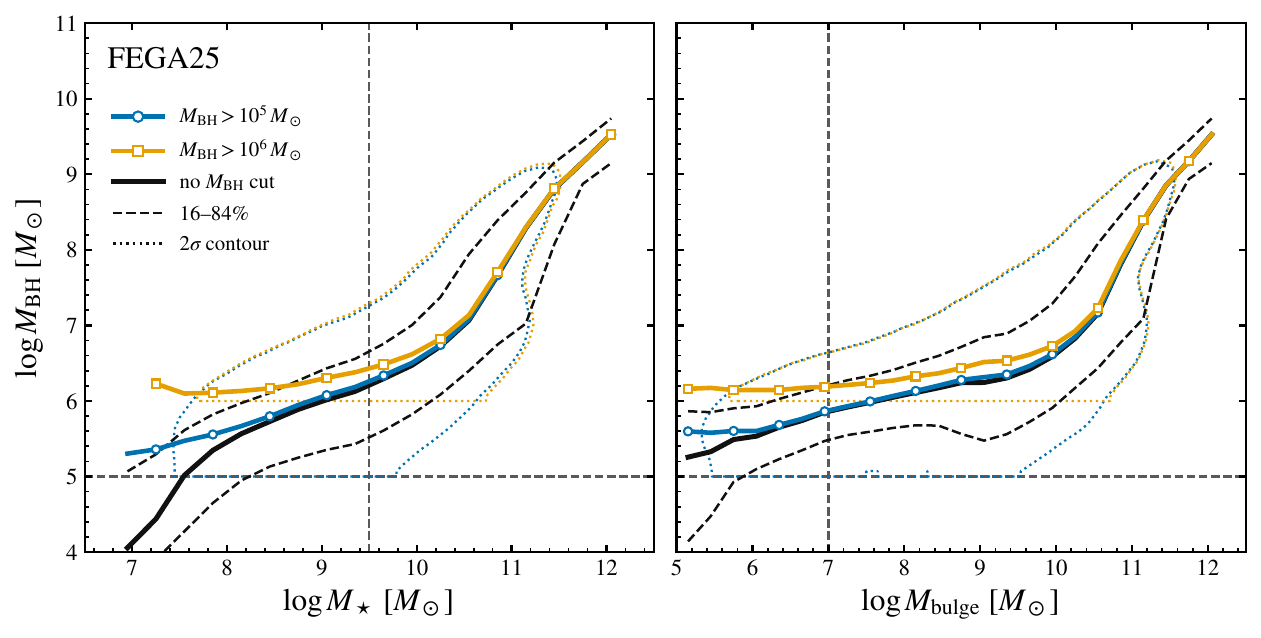}
\caption{Black hole--host galaxy scaling relations predicted by \texttt{FEGA25} at $z=0$. The left panel shows $M_{\rm BH}$ as a function of total stellar mass, while the right panel shows the relation with bulge mass. Black solid curves indicate the median relations obtained without imposing a black hole mass cut, with black dashed curves enclosing the 16th--84th percentile range. Blue circles and orange squares show the median relations obtained after selecting black holes with $M_{\rm BH}>10^{5}\,{\rm M}_{\odot}$ and $M_{\rm BH}>10^{6}\,{\rm M}_{\odot}$, respectively, while the corresponding dotted contours enclose the $2\sigma$ regions of the selected distributions. The horizontal dashed line marks $M_{\rm BH}=10^{5}\,{\rm M}_{\odot}$. The vertical dashed lines indicate $\log(M_{\star}/{\rm M}_{\odot})=9.5$ in the left panel and $\log(M_{\rm bulge}/{\rm M}_{\odot})=7$ in the right panel. The comparison illustrates how imposing a black hole mass threshold preferentially affects the low-mass regime, whereas the selected and no-cut relations progressively converge toward higher host masses.}
\label{fig:fig1}
\end{figure*}

\begin{figure*}[t!]
\centering
\includegraphics[width=0.95\textwidth]{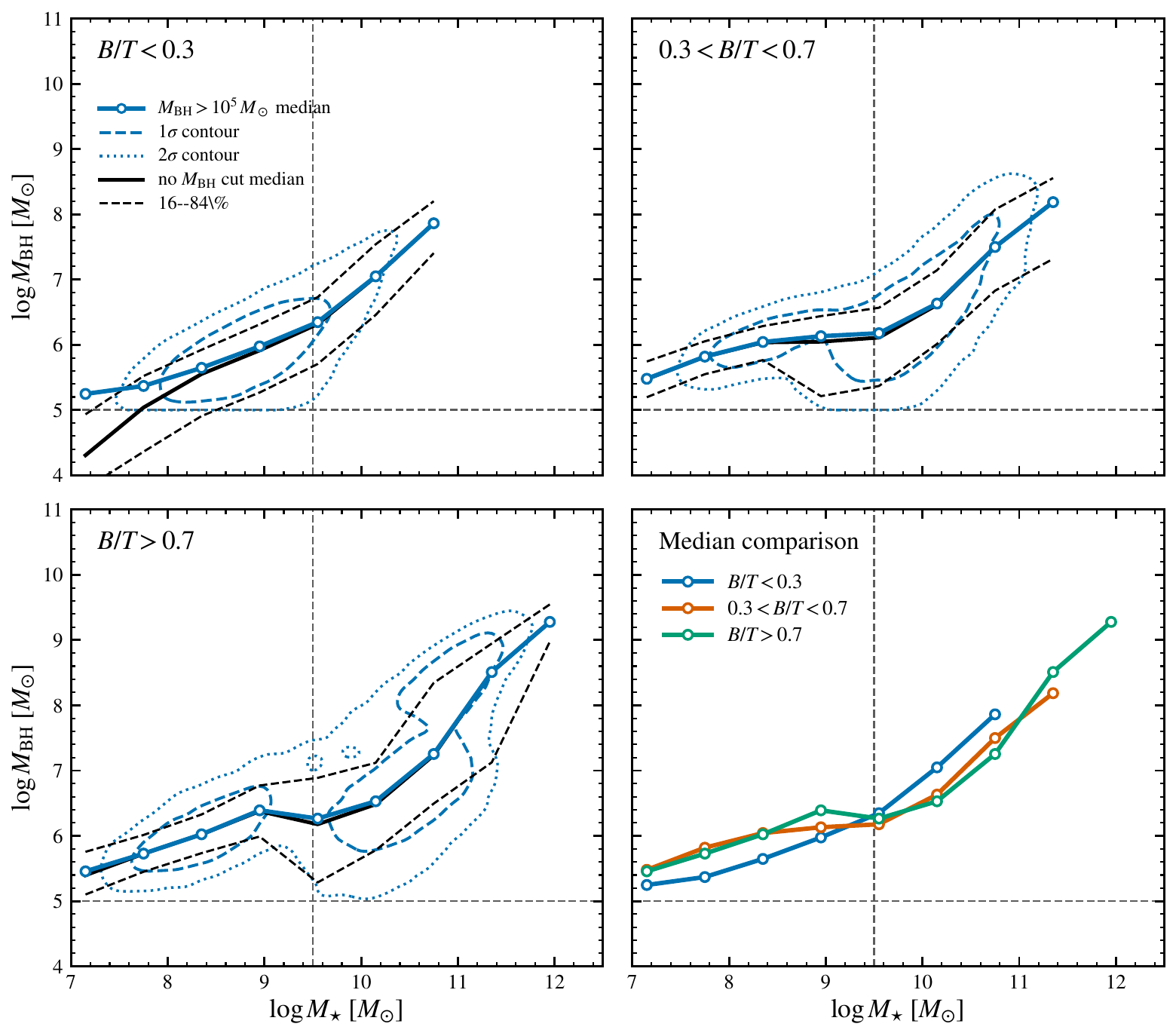}
\caption{Black hole mass as a function of total stellar mass predicted by FEGA25 at $z=0$, separated according to galaxy morphology, characterized by the bulge-to-total stellar mass ratio ($B/T$). The first three panels show the relations for the three morphological classes, while the fourth panel provides a direct comparison among them. Black solid curves indicate the median relations obtained without imposing a black hole mass cut, with black dashed curves enclosing the corresponding 16th--84th percentile ranges. Blue solid curves show the median relations for galaxies hosting black holes with $M_{\rm BH}>10^5\,M_\odot$, while blue dashed and dotted curves indicate the corresponding $1\sigma$ and $2\sigma$ regions, respectively.}
\label{fig:fig2}
\end{figure*}

In this section, we focus on the relations between BH mass and total stellar or bulge mass, and on their possible dependence on galaxy morphology, quantified through the bulge-to-total stellar mass ratio, $B/T$. We consider two different BH mass thresholds, $M_{\rm BH}>10^{5}\,{\rm M}_{\odot}$ and $M_{\rm BH}>10^{6}\,{\rm M}_{\odot}$, together with the intrinsic relations obtained without imposing any BH mass cut. \footnote{Unlike \texttt{NewCluster}, \texttt{FEGA25} does not impose a fixed BH seed mass. Instead, BHs are initially formed through cold-gas accretion during merger-driven quasar-mode events, with their initial masses determined by the accreted gas rather than by a prescribed seed value.}

The reference lines in Figure~\ref{fig:fig1} are approximate guides to the eye rather than strict completeness limits.

Figure~\ref{fig:fig1} clearly shows that the BH mass threshold imposed on the sample mainly affects the low-mass regime, whereas the selected relations progressively converge toward the no-cut relation at higher host masses. This behaviour is observed for both total stellar mass and bulge mass. In particular, the relation obtained with the lowest BH mass threshold is nearly indistinguishable from the no-cut relation over most of the mass range explored. By contrast, the more restrictive $M_{\rm BH}>10^{6}\,{\rm M}_{\odot}$ selection produces appreciable deviations at the low-mass end.

The close agreement between the relations obtained with $M_{\rm BH}>10^{5}\,M_\odot$ and those derived without imposing any BH mass selection suggests that $10^{5}\,M_\odot$ represents a suitable fiducial threshold for our analysis. At this limit, the selected sample preserves the underlying BH--host relation over nearly the entire host-mass range explored, while still providing a well-defined BH population for comparison with other theoretical and observational studies. A BH mass scale of $\sim10^{5}\,M_\odot$ is also commonly encountered in numerical models, either as a characteristic seed mass or as a lower threshold adopted in studies of BH demographics \citep{dubois2014,schaye2015,sijacki2015,han2026,bhowmick2025,bhowmick2026}. Observationally, samples specifically targeting low-mass active BHs extend into the same mass regime, with BH masses of $\sim10^{5}$--$10^{6}\,M_\odot$ and, in some cases, even below $10^{5}\,M_\odot$ \citep{jiang2011,baldassare2015,liu2018,greene2020}. We therefore adopt $M_{\rm BH}>10^{5}\,M_\odot$ as our reference BH mass selection, while the higher threshold of $10^{6}\,M_\odot$ is used to quantify the impact of a progressively more restrictive selection.

Figure~\ref{fig:fig2} separates the sample into three morphological bins ($B/T<0.3$, $0.3<B/T<0.7$, and $B/T>0.7$), with the fourth panel directly comparing their median $M_{\rm BH}$--$M_\star$ relations.

The median relations are broadly comparable over their overlapping stellar-mass ranges, particularly at low and intermediate masses, while the bulge-dominated population extends to higher stellar masses. The distributions around the median relations also vary with morphology.

The morphology dependence suggests that part of the scatter in the global $M_{\rm BH}$--$M_\star$ relation may reflect the superposition of galaxies with different structural properties and assembly histories, consistent with the observational evidence discussed in the Introduction.

\subsection{Redshift Dependence}
\label{sec:redshift}

\begin{figure*}[t!]
\centering
\includegraphics[width=0.94\textwidth]{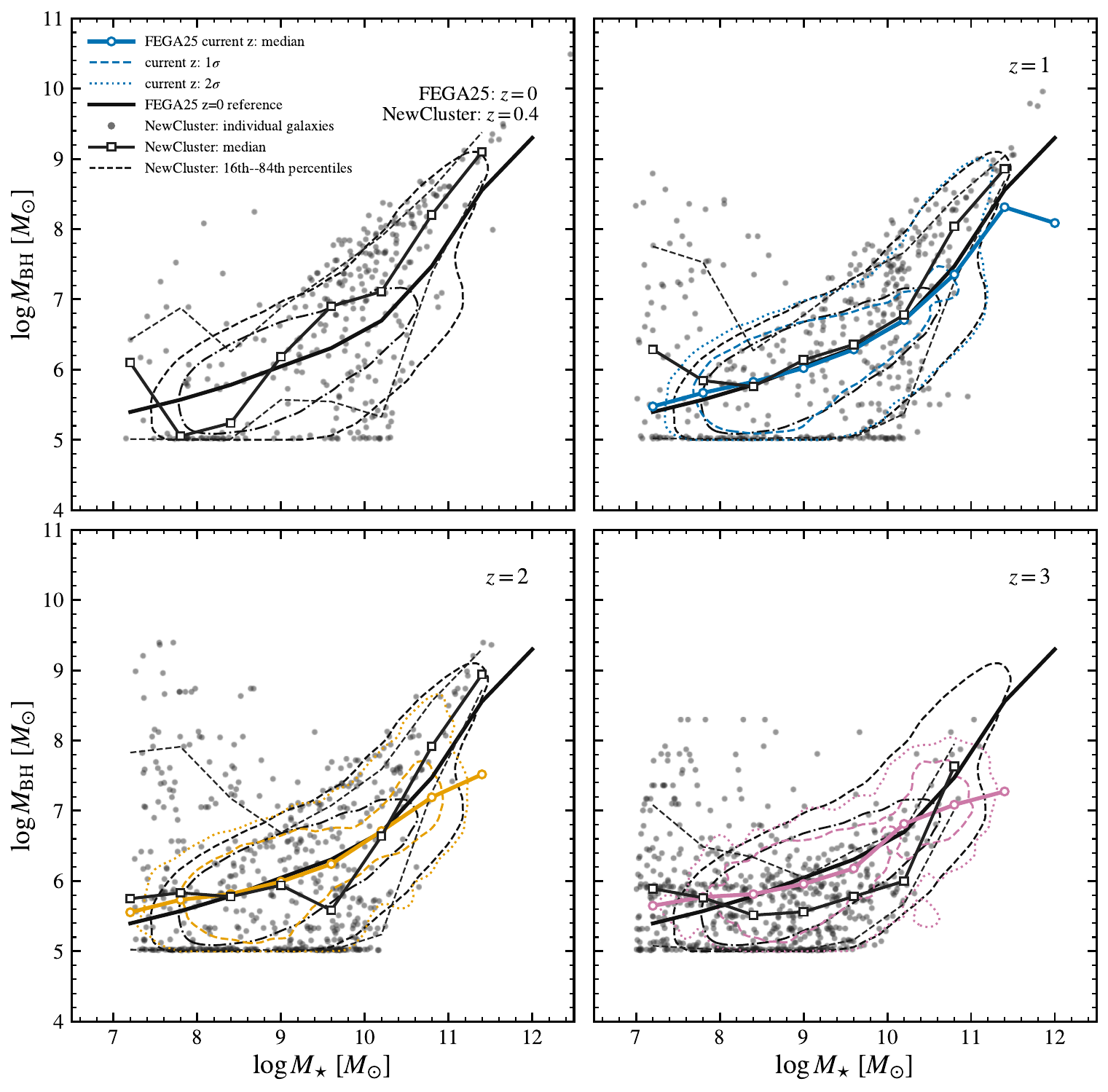}
\caption{Evolution of the $M_{\rm BH}$--$M_{\star}$ relation predicted by \texttt{FEGA25} and comparison with the hydrodynamical zoom-in simulation \texttt{NewCluster}. The black solid line shows the median relation predicted by \texttt{FEGA25} at $z=0$ and is repeated in all panels as a reference. At $z=1$, 2, and 3, the colored solid lines with open circles show the corresponding \texttt{FEGA25} median relations, while the associated contours enclose the $1\sigma$ and $2\sigma$ regions. Dark-gray points show individual galaxies from \texttt{NewCluster}; the solid line with open squares indicates their median relation, and the dashed lines show the 16th and 84th percentiles. In the first panel, the \texttt{FEGA25} prediction at $z=0$ is compared with \texttt{NewCluster} at $z=0.4$, the lowest redshift available for the hydrodynamical simulation. Only BHs with $M_{\rm BH}\geq10^{5}\,M_{\odot}$ are considered. Contours extending below this BH mass limit are displayed at $\log(M_{\rm BH}/M_{\odot})=5$.}
\label{fig:fig3}
\end{figure*}

\begin{figure*}[t!]
\centering
\includegraphics[width=0.94\textwidth]{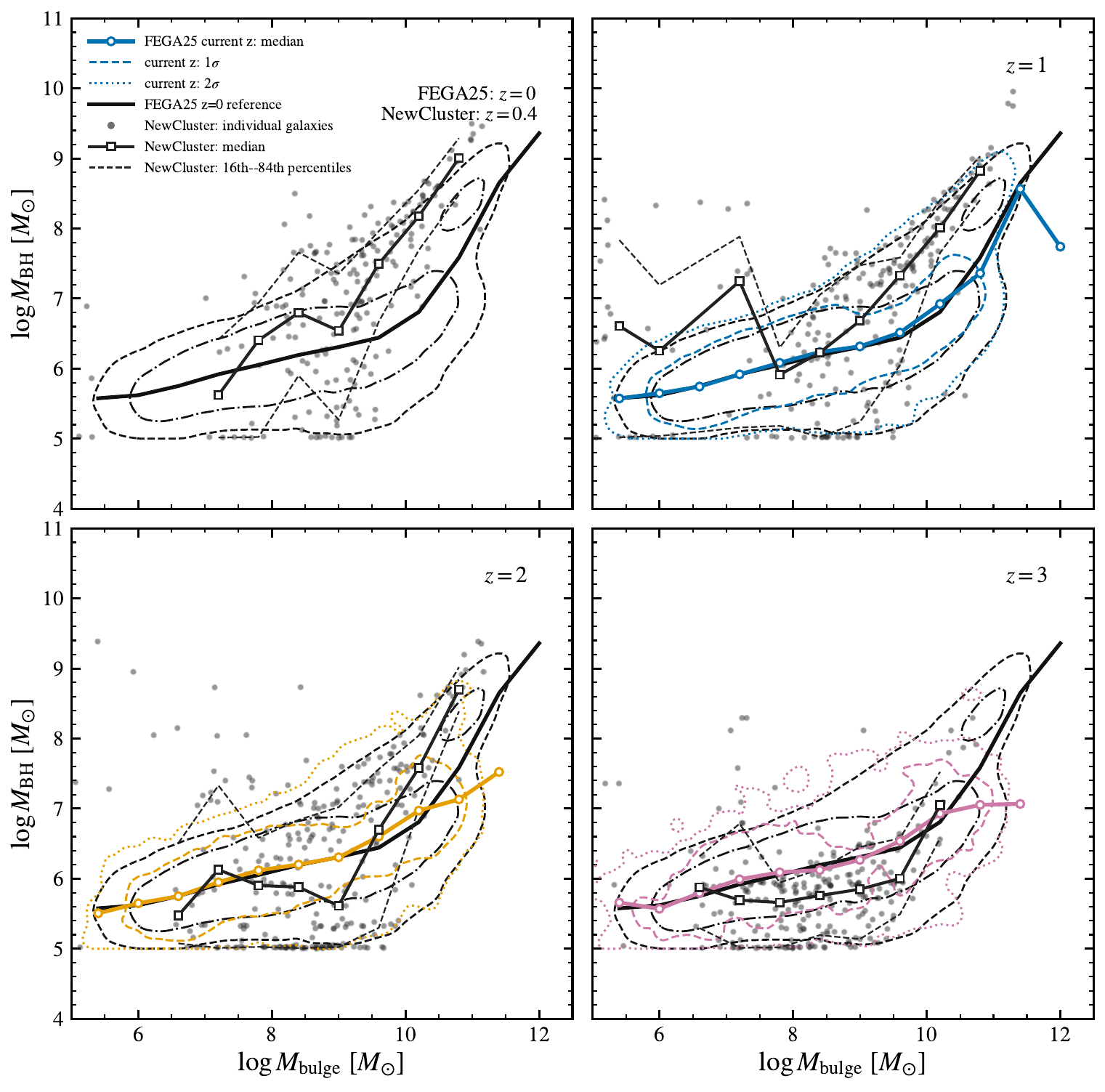}
\caption{ Evolution of the $M_{\rm BH}$--$M_{\rm bulge}$ relation predicted by \texttt{FEGA25} and comparison with the hydrodynamical zoom-in simulation \texttt{NewCluster}. The black solid line shows the median relation predicted by \texttt{FEGA25} at $z=0$ and is repeated in all panels as a reference. At $z=1$, 2, and 3, the colored solid lines with open circles show the corresponding \texttt{FEGA25} median relations, while the associated contours enclose the $1\sigma$ and $2\sigma$ regions. Dark-gray points show individual galaxies from \texttt{NewCluster}; the solid line with open squares indicates their median relation, and the dashed lines show the 16th and 84th percentiles. In the first panel, the \texttt{FEGA25} prediction at $z=0$ is compared with \texttt{NewCluster} at $z=0.4$. Only BHs with $M_{\rm BH}\geq10^{5}\,M_{\odot}$ are considered. Contours extending below this BH mass limit are displayed at $\log(M_{\rm BH}/M_{\odot})=5$.}
\label{fig:fig4}
\end{figure*}

Figures~\ref{fig:fig3} and~\ref{fig:fig4} show the evolution of the $M_{\rm BH}$--$M_{\star}$ and $M_{\rm BH}$--$M_{\rm bulge}$ relations, respectively. In each panel, the $z=0$ relation predicted by \texttt{FEGA25} is shown as a reference, while the corresponding relations at higher redshift are reported at $z=1$, 2, and 3. Overall, the model predicts no strong systematic evolution of either scaling relation over the redshift range considered. At low and intermediate host masses, the median relations at different redshifts remain remarkably similar, with substantial overlap between their distributions. Larger deviations appear toward the highest host masses, particularly at $z\gtrsim2$, although these regimes are also those most affected by the decreasing number of massive systems. Most importantly, the differences do not describe a clear monotonic shift with redshift, suggesting that the normalization of the relations remains broadly stable across cosmic time.

The weak redshift dependence predicted by FEGA25 is consistent with a scenario in which BHs and their host galaxies undergo a statistically coupled growth, preserving approximately the same BH--host mass relations throughout cosmic time. This suggests that the connection between BH growth and galaxy assembly may already be established at $z\sim3$, with subsequent evolution proceeding without substantially altering the underlying scaling relations. However, this result does not necessarily imply synchronized growth of individual BHs and their hosts, as the relations describe galaxy populations at different redshifts rather than the evolutionary trajectories of individual systems.

The \texttt{NewCluster} simulation provides an independent hydrodynamical comparison. At $z=1$, the median $M_{\rm BH}$--$M_{\star}$ relation is broadly consistent with that predicted by \texttt{FEGA25} over a large fraction of the common stellar mass range. At higher redshift, the \texttt{NewCluster} distributions show a larger scatter and a more pronounced variation toward the most massive galaxies, while the median relation at low and intermediate masses remains comparatively stable. A qualitatively similar behaviour is found for the $M_{\rm BH}$--$M_{\rm bulge}$ relation, although the scatter and the differences between the two models are somewhat more pronounced. In both cases, \texttt{NewCluster} does not show a simple monotonic displacement of the relation with increasing redshift.

The observational picture of the redshift evolution of BH--host relations remains debated. Several studies have reported an increase of the BH-to-host mass ratio with increasing redshift
\citep{peng2006,merloni2010,bennert2011,ding2020}, whereas other works have found little or no significant evolution \citep{jahnke2009,cisternas2011,suh2020}. In addition, the inferred
evolution can be strongly affected by the selection of active BHs and their host galaxies. In particular, accounting for these effects can substantially reduce the apparent offset from the local relations \citep{schulze2014}, consistent with the broader evidence that BH scaling relations are particularly sensitive to sample selection \citep{lauer2007,schulze2011,shankar2016,shankar2019}. The absence of significant evolution predicted by \texttt{FEGA25} should therefore be regarded as one of the scenarios currently allowed by the observational constraints rather than as a uniquely established evolutionary picture.

The comparison with \texttt{NewCluster} should nevertheless be interpreted with caution. \texttt{NewCluster} follows a highly overdense cluster region and does not sample a representative
cosmological volume, while \texttt{FEGA25} is applied to a large cosmological box. Differences between the two predictions can therefore reflect not only the different numerical and physical modelling, but also the distinct environments, assembly histories, and galaxy populations sampled by the two simulations. Moreover, the lowest available redshift for \texttt{NewCluster} is $z=0.4$, rather than $z=0$. We therefore regard the hydrodynamical simulation primarily as a qualitative, independent comparison. Taken together, the two approaches indicate that the overall form of the BH--host relations is already present at high redshift and undergoes no strong systematic evolution over the range explored here, although their scatter and high-mass behaviour can depend substantially on the galaxy population and environment considered.

\subsection{Connecting Black Hole Occupation Fraction and Black Hole--Host Scaling Relations}
\label{sec:connection}

\begin{figure*}[t!]
\centering
\includegraphics[width=0.94\textwidth]{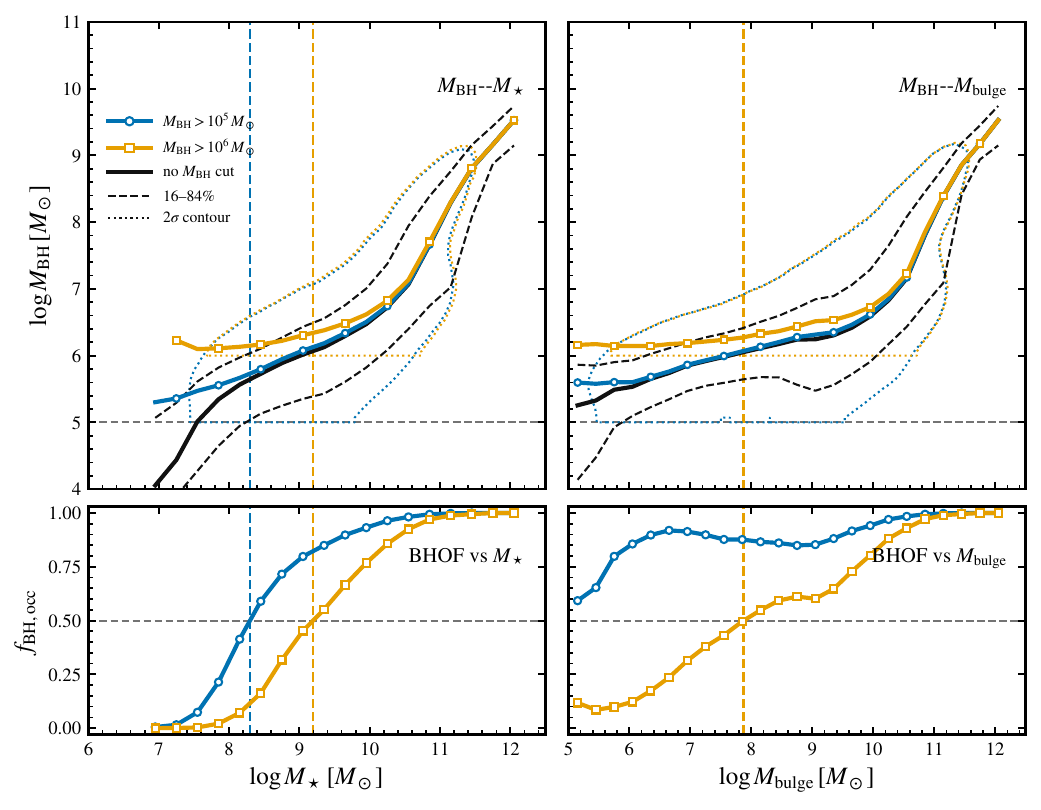}
\caption{Connection between black hole--host scaling relations and black hole occupation fractions predicted by \texttt{FEGA25} at $z=0$. The upper panels show $M_{\rm BH}$ as a function of total stellar mass (left) and bulge stellar mass (right), while the lower panels show the corresponding black hole occupation fractions, $f_{\rm BH,occ}$. Blue circles and orange squares refer to the selections $M_{\rm BH}>10^{5}\,{\rm M}_{\odot}$ and $M_{\rm BH}>10^{6}\,{\rm M}_{\odot}$, respectively, while the black curves in the upper panels show the no-cut relations. The vertical dashed lines mark the host masses at which $f_{\rm BH,occ}=0.5$, determined by interpolation between adjacent mass bins. For the stellar-mass relation, these occur at $\log(M_{\star,50}/{\rm M}_{\odot})\simeq8.30$ and $9.19$ for the $10^{5}$ and $10^{6}\,{\rm M}_{\odot}$ thresholds, respectively. For the bulge-mass relation, the $10^{6}\,{\rm M}_{\odot}$ threshold reaches $f_{\rm BH,occ}=0.5$ at $\log(M_{{\rm bulge},50}/{\rm M}_{\odot})\simeq7.88$, whereas the occupation fraction for $M_{\rm BH}>10^{5}\,{\rm M}_{\odot}$ is already above 0.5 at the lower edge of the plotted bulge-mass range. The figure shows that large differences in occupation fraction do not necessarily translate into equally large differences in the corresponding median scaling relations, highlighting the complementary information carried by black hole demographics and black hole--host scaling relations.}
\label{fig:fig5}
\end{figure*}

\begin{figure*}[t!]
\centering
\includegraphics[width=0.94\textwidth]{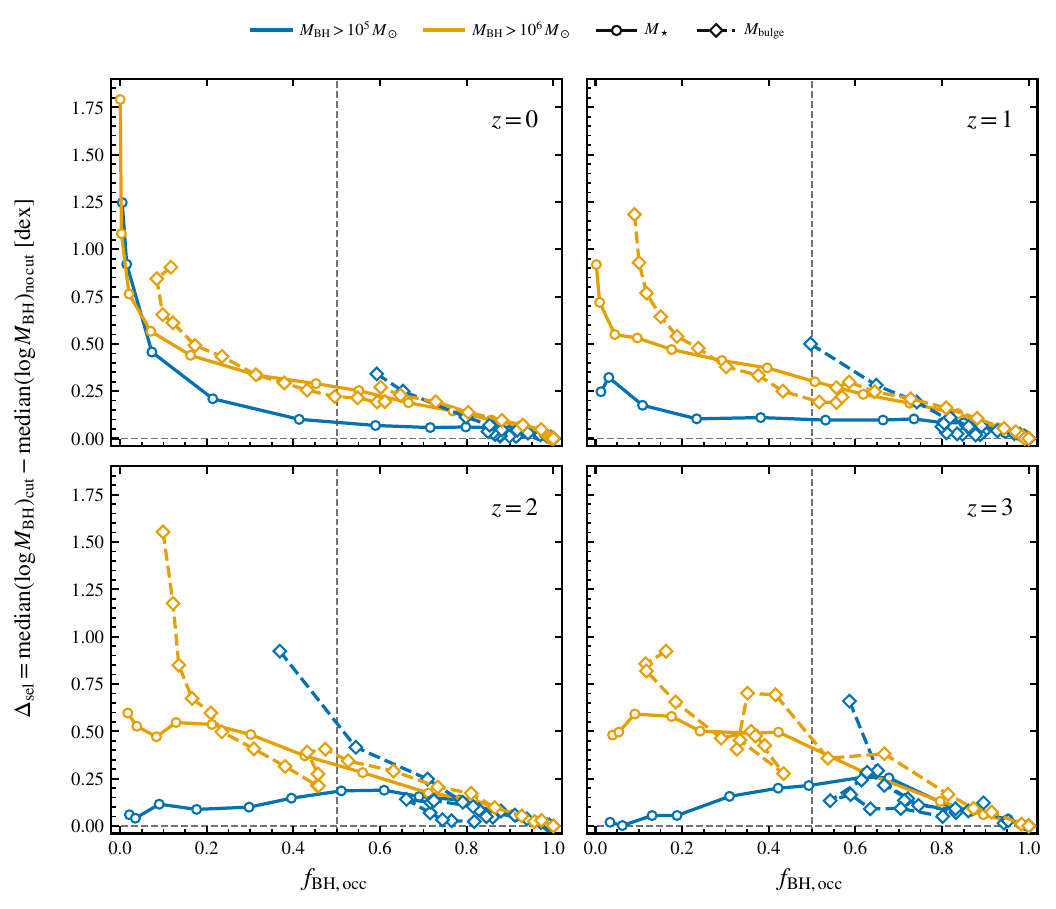}
\caption{Selection-induced offset in the black hole--host scaling relations as a function of the black hole occupation fraction predicted by \texttt{FEGA25}. The four panels show the results at $z=0$, 1, 2, and 3. The vertical axis is defined as $\Delta_{\rm sel}={\rm median}(\log M_{\rm BH})_{\rm cut}-{\rm median}(\log M_{\rm BH})_{\rm no\,cut}$ and therefore quantifies, in each host-mass bin, the upward shift of the median scaling relation induced by the adopted black hole mass threshold. Blue and orange curves correspond to $M_{\rm BH}>10^{5}\,{\rm M}_{\odot}$ and $M_{\rm BH}>10^{6}\,{\rm M}_{\odot}$, respectively. Solid curves with open circles show the results obtained when binning galaxies by total stellar mass, while dashed curves with open diamonds refer to bulge stellar mass. The vertical dashed line marks $f_{\rm BH,occ}=0.5$, and the horizontal dashed line indicates $\Delta_{\rm sel}=0$. Large values of $\Delta_{\rm sel}$ are preferentially associated with low occupation fractions, where the black hole mass cut selects only the high-$M_{\rm BH}$ tail of the underlying population. As $f_{\rm BH,occ}$ approaches unity, the selected and no-cut median relations progressively converge and $\Delta_{\rm sel}$ approaches zero. At fixed $f_{\rm BH,occ}$, however, the offset can differ substantially between the stellar-mass and bulge-mass relations, showing that the occupation fraction alone does not uniquely determine the selection bias and that its impact also depends on the host property used to define the scaling relation.}
\label{fig:fig6}
\end{figure*}

In this section, we connect the BH--host scaling relations with the corresponding BH occupation fraction and quantify how BH mass selection biases the inferred relations. We first consider the behaviour at $z=0$ and then examine how the relation between BHOF and selection bias evolves with redshift.

Figure~\ref{fig:fig5} compares the $z=0$ BHOFs with the corresponding selected and no-cut BH--host relations.

The figure directly links the host mass at which a given BH population becomes common to the effect of the corresponding mass selection on the scaling relation. In terms of total stellar mass, the occupation fraction reaches $f_{\rm BH,occ}=0.5$ at $\log(M_{\star,50}/{\rm M}_\odot)\simeq8.30$ and $9.19$ for the $10^{5}$ and $10^{6}\,{\rm M}_\odot$ thresholds, respectively. Thus, increasing the BH threshold by one dex shifts the stellar mass at which half of the galaxies host a BH above the adopted threshold by approximately $0.9$ dex.

Below these characteristic masses, the selected relations preferentially sample the high-$M_{\rm BH}$ tail and therefore lie above the no-cut relation. As the BHOF approaches unity, the selected relations progressively converge toward the intrinsic relation. Such a sensitivity of BH--host scaling relations to sample selection has also been emphasized in both numerical and observational studies \citep{degraf2015,shankar2016,shankar2019}. Importantly, however, a substantial difference in BHOF does not necessarily produce a similarly large difference in the median scaling relation. For example, at $\log(M_\star/{\rm M}_\odot)\simeq9.05$, the BHOFs are approximately $0.80$ and $0.45$ for the two thresholds, whereas the corresponding median BH masses differ by only $\simeq0.23$ dex.

The bulge-mass dependence shows a somewhat different behaviour. For $M_{\rm BH}>10^{5}\,{\rm M}_\odot$, the BHOF among galaxies with $M_{\rm bulge}>0$ is already above $0.5$ at the lower boundary of the plotted range, $\log(M_{\rm bulge}/{\rm M}_\odot)=5$, and the corresponding scaling relation remains close to the no-cut case over most of the mass range. For the $10^{6}\,{\rm M}_\odot$ threshold, instead, $f_{\rm BH,occ}=0.5$ is reached at $\log(M_{{\rm bulge},50}/{\rm M}_\odot)\simeq7.88$. Below this value the selection has a stronger effect, whereas at higher bulge masses the increasing BHOF is accompanied by convergence toward the no-cut relation. As in the stellar-mass case, even substantial differences in occupation can correspond to relatively small differences between the selected median relations.

These results show that a BH mass threshold primarily changes the fraction of the galaxy population represented in the selected sample, while its effect on the median BH--host relation can be considerably smaller. Similar scaling relations therefore do not necessarily imply similar underlying BH populations. In this respect, the BHOF provides information that cannot be inferred from the scaling relation alone, particularly toward low galaxy masses where BH occupation remains an important constraint on BH demographics and formation scenarios \citep{miller2015,greene2020,haidar2022,contini2026BH}.

The comparison between stellar and bulge mass further suggests that the two host properties encode BH occupation differently. The transition from low to high occupation is clearly visible as a function of total stellar mass, whereas the BHOF as a function of bulge mass is already relatively high at low $M_{\rm bulge}$, particularly for the $10^{5}\,{\rm M}_\odot$ threshold. In \texttt{FEGA25}, this behaviour reflects the closer connection between bulge growth and BH formation, although this interpretation must be treated with care because BH formation itself is linked to the presence of a bulge in the model.

Figure~\ref{fig:fig6} further investigates the connection between BH occupation and sample selection by showing the BH occupation fraction, $f_{\rm BH,occ}$, against the selection-induced offset in the median BH--host scaling relation,
\begin{equation}
\Delta_{\rm sel} =
{\rm median}\left[\log M_{\rm BH}\right]_{\rm cut}
-
{\rm median}\left[\log M_{\rm BH}\right]_{\rm no\,cut},
\end{equation}
for both stellar mass and bulge mass at $z=0$, 1, 2, and 3. The two BH mass thresholds, $M_{\rm BH}>10^{5}\,{\rm M}_{\odot}$ and $M_{\rm BH}>10^{6}\,{\rm M}_{\odot}$, are shown separately.

A clear general trend is present at all redshifts: large values of $\Delta_{\rm sel}$ are preferentially associated with low BHOFs, whereas $\Delta_{\rm sel}$ tends toward zero as $f_{\rm BH,occ}$ approaches unity. When the occupation fraction is low, only a small fraction of galaxies in a given host-mass bin satisfy the adopted BH mass threshold, and the selected sample preferentially traces the high-$M_{\rm BH}$ tail of the underlying distribution. The resulting median BH mass is therefore shifted upward relative to the no-cut relation. As the BHOF increases, the selected population becomes progressively more representative and the offset decreases. This behaviour is consistent with the general expectation that selection in BH mass can substantially bias BH--host scaling relations, particularly toward the low-mass regime \citep{lauer2007,schulze2011,degraf2015,shankar2016,shankar2019}.

The relation between $f_{\rm BH,occ}$ and $\Delta_{\rm sel}$ is nevertheless not one-to-one. At fixed occupation fraction, the $M_{\rm BH}$--$M_{\star}$ and $M_{\rm BH}$--$M_{\rm bulge}$ relations can exhibit substantially different offsets. This occurs because the two quantities encode different aspects of the BH population: the BHOF measures the fraction of galaxies above the adopted BH threshold, whereas $\Delta_{\rm sel}$ measures how strongly that selection changes the median BH mass. Two galaxy samples can therefore have the same BHOF but different values of $\Delta_{\rm sel}$ if their underlying BH mass distributions differ.

This effect is particularly evident at $z=0$--2, where the bulge-based relation frequently shows a larger $\Delta_{\rm sel}$ than the stellar-mass relation at comparable occupation fractions. Thus, although the largest absolute offsets can occur in the stellar-mass relation at very low BHOF, this does not imply that it is systematically more affected by selection. At $z=3$, the distinction between the two host properties becomes non-linear, and their relative sensitivity depends more strongly on both BHOF and the adopted BH mass threshold.

Figure~\ref{fig:fig6} therefore demonstrates that there is no universal mapping between BHOF and selection bias. A low BHOF generally signals that a scaling relation may be strongly affected by the imposed BH mass threshold, but it does not uniquely determine the magnitude of that bias. At fixed $f_{\rm BH,occ}$, $\Delta_{\rm sel}$ depends on the host property used to construct the relation and on the underlying BH mass distribution, with this dependence also evolving with redshift.

Together with Figure~\ref{fig:fig5}, this result reinforces the complementary nature of BHOF and BH--host scaling relations. The former quantifies how much of the underlying galaxy population is represented above a given BH mass threshold, while the latter characterizes the BHs that survive that selection. Consequently, similar occupation fractions do not necessarily imply similar selection biases, just as similar scaling relations do not necessarily imply similarly sampled underlying BH populations.

\subsection{Comparison With Observations}
\label{sec:comparison}

\begin{figure*}[t!]
\centering
\includegraphics[width=0.94\textwidth]{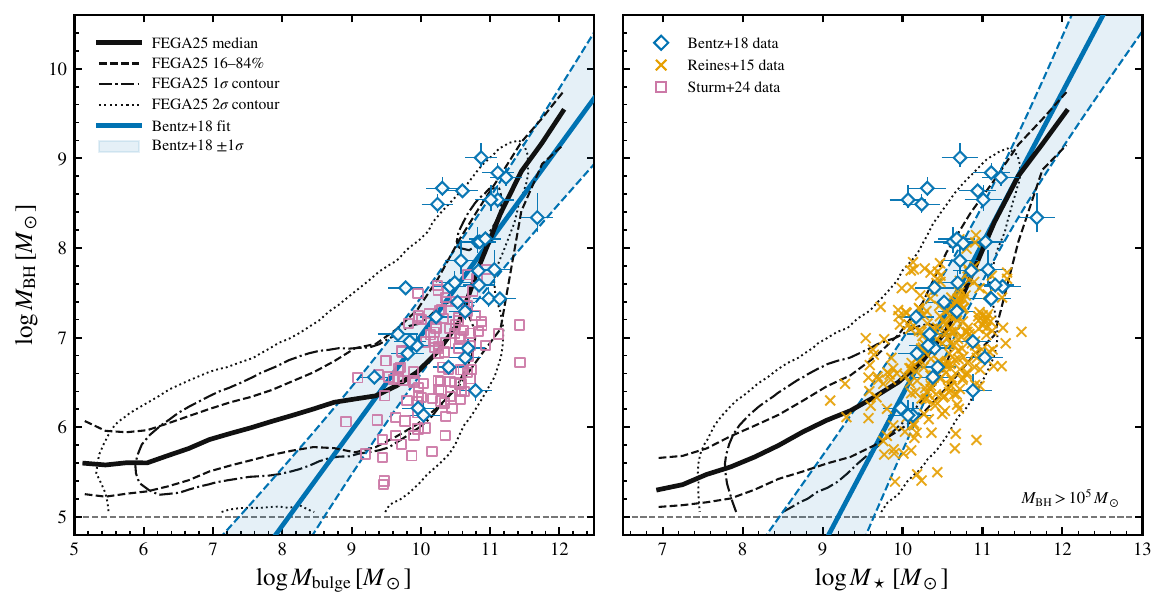}
\caption{Comparison between the black hole--host galaxy scaling relations predicted by \texttt{FEGA25} at $z=0$ and local observational samples. The left panel shows black hole mass as a function of bulge stellar mass, while the right panel shows the relation with total stellar mass. The \texttt{FEGA25} predictions are obtained by selecting black holes with $M_{\rm BH}>10^{5}\,{\rm M}_{\odot}$. Black solid curves show the median relations, black dashed curves indicate the 16th--84th percentile range, and black dash-dotted and dotted contours enclose the $1\sigma$ and $2\sigma$ regions, respectively. The density contours are restricted to $\log(M_{\rm BH}/{\rm M}_{\odot})\geq5$. Blue solid curves show the best-fitting relations derived by \citet{bentz2018}, with the corresponding $\pm1\sigma$ regions indicated by the blue shaded areas and dashed boundaries. Blue open diamonds show the individual measurements from \citet{bentz2018}. Magenta open squares in the left panel show the bulge-mass measurements from \citet{sturm2024}, while orange crosses in the right panel show the total stellar-mass measurements from \citet{reines2015}. The comparison illustrates the agreement between the \texttt{FEGA25} predictions and the observed local black hole--host galaxy relations over the mass range probed by the data, while highlighting the substantial scatter toward lower host masses.}
\label{fig:fig7}
\end{figure*}

We conclude our analysis by comparing the predictions of \texttt{FEGA25} with observational measurements. Figure~\ref{fig:fig7} shows the $M_{\rm BH}$--$M_{\rm bulge}$ relation in the left panel and the $M_{\rm BH}$--$M_{\star}$ relation in the right panel. The model predictions are obtained by selecting BHs with $M_{\rm BH}>10^{5}\,{\rm M}_{\odot}$. Black solid curves show the median relations, black dashed curves indicate
the 16th--84th percentile range, and black dash-dotted and dotted contours enclose the $1\sigma$ and $2\sigma$ regions, respectively. The density contours are restricted to $\log(M_{\rm BH}/{\rm M}_{\odot})\geq5$.

We compare the model predictions with the best-fitting relations from \citet{bentz2018}, shown by the blue solid curves, with the corresponding $\pm1\sigma$ regions indicated by the blue shaded areas and dashed
boundaries. Blue open diamonds show the individual measurements from \citet{bentz2018}. Magenta open squares in the left panel represent the bulge-mass measurements from \citet{sturm2024}, while orange crosses in the
right panel show the total stellar-mass measurements from \citet{reines2015}.

Overall, the \texttt{FEGA25} predictions are in good agreement with the available observational constraints for both scaling relations. Most of the observed galaxies lie within the $2\sigma$ regions predicted by the model, although the observational coverage becomes increasingly sparse toward lower masses. In this poorly sampled regime, \texttt{FEGA25} predicts a bending of both relations toward lower host masses, which cannot yet be robustly tested with the data considered here. Both the observations and the model also show substantial scatter. In \texttt{FEGA25}, this scatter becomes particularly large toward the low-mass regime, where the predicted BH population spans a broad range of masses at fixed host mass.

The agreement at $z=0$ is particularly relevant when considered together with the absence of significant redshift evolution predicted by \texttt{FEGA25}. As discussed in Section~\ref{sec:redshift}, the observational picture remains debated and is strongly affected by sample-selection effects. The local agreement found here therefore provides an important reference point for interpreting the predicted evolution of the scaling relations, while emphasizing that an apparent offset from the local relation at higher redshift does not necessarily imply intrinsic evolution of the underlying BH--host connection.

\section{Conclusions}
\label{sec:conclusion}

We take advantage of the \texttt{FEGA25} semianalytic model of galaxy formation and the hydrodynamical zoom-in simulation \texttt{NewCluster} to investigate the $M_{\rm BH}$--$M_{\star}$ and $M_{\rm BH}$--$M_{\rm bulge}$ relations, their dependence on BH mass threshold, galaxy morphology, and redshift, and their connection with the BH occupation fraction. The main conclusions of our analysis can be summarized as follows.

\begin{itemize}

\item The $M_{\rm BH}$--$M_{\star}$ and $M_{\rm BH}$--$M_{\rm bulge}$ relations are sensitive to the adopted BH mass threshold primarily at low host masses, while progressively converging toward the intrinsic no-cut
relations at higher masses. The $M_{\rm BH}$--$M_{\star}$ relation also shows a clear dependence on galaxy morphology, quantified here through the bulge-to-total stellar mass ratio. Increasingly bulge-dominated systems preferentially populate higher stellar masses. The three morphological classes occupy different stellar-mass ranges, while their median relations are broadly similar where they overlap.

\item We find no significant evolution of the BH--host scaling relations with redshift over the range explored. This behaviour is consistent with a picture of BH--galaxy coevolution in which the statistical connection between BH mass and host-galaxy properties is established relatively early and remains broadly preserved during subsequent cosmic evolution. The independent comparison with \texttt{NewCluster} supports a qualitatively similar picture, showing no clear monotonic evolution, although the hydrodynamical simulation exhibits substantial scatter and larger differences toward the high-mass regime. Since \texttt{NewCluster} samples a highly overdense cluster environment, this comparison should be interpreted primarily as a qualitative consistency check rather than as a direct comparison between statistically equivalent galaxy populations. The observational picture remains debated: several studies report little or no evolution, whereas others find increasing BH-to-host mass ratios toward higher redshift. Our predictions are therefore consistent with one of the scenarios currently allowed by the observations, while emphasizing the importance of accounting for sample-selection effects when interpreting apparent redshift evolution.

\item BH occupation fraction and BH--host scaling relations provide complementary information on the underlying BH population. Although a low BHOF generally corresponds to a stronger selection-induced offset, the relation between occupation fraction and bias is not unique: substantially different BHOFs can produce very similar median scaling relations, while at fixed BHOF the magnitude of the selection bias depends on the host property considered and on the underlying BH mass distribution. Thus, the representativeness of a BH-selected sample cannot be inferred from the scaling relation alone, nor can the impact of BH mass selection be
determined from the BHOF alone; both quantities must be considered jointly.

\item At $z=0$, \texttt{FEGA25} reproduces the observed $M_{\rm BH}$--$M_{\star}$ and $M_{\rm BH}$--$M_{\rm bulge}$ relations over the mass range currently constrained by the data, including their substantial scatter. This agreement shows that the BH growth and galaxy-evolution prescriptions implemented in the model are consistent with the available local observational constraints, while the predicted bending and increasing scatter toward lower host masses provide testable predictions for future observational samples.

\end{itemize}

Overall, our results highlight the importance of considering BH--host scaling relations together with the demographic information encoded by the BHOF. While scaling relations describe the typical masses of BHs hosted by galaxies with given stellar or bulge masses, the BHOF provides complementary information on how representative the selected BH population is of the underlying galaxy population. Our analysis demonstrates that these two quantities are not uniquely connected: similar scaling relations can arise from substantially different occupation fractions, while comparable BHOF values do not necessarily imply similar selection-induced biases. This distinction becomes particularly important in the low-mass regime, where BH mass thresholds can strongly affect the inferred relations and where uncertainties in BH occupation remain significant.

The dependence of the scaling relations on galaxy morphology further emphasizes that the connection between BH growth and galaxy evolution cannot be fully characterized by a single, universal relation. A joint analysis of scaling relations, morphological properties, and occupation fractions therefore offers a more comprehensive framework for interpreting the observed BH population and for testing the physical prescriptions implemented in theoretical models. In particular, combining these complementary diagnostics may help distinguish between different scenarios of BH formation and subsequent growth, although degeneracies between these processes are expected to remain.

Looking ahead, this approach will become increasingly valuable as future observational surveys extend reliable BH mass measurements toward lower host masses and higher redshifts. Accounting for observational selection effects and BH mass completeness will be essential for meaningful comparisons between observations and theoretical predictions. Ultimately, the combined study of BH--host scaling relations and BH occupation fractions promises to provide stronger constraints on the physical processes governing the coevolution of galaxies and their central BHs across cosmic time.

\section*{Acknowledgements}
E.C. and S.K.Y. acknowledge support from the Korean National Research Foundation (RS-2025-00514475). E.C. acknowledges support from the Korean National Research Foundation (RS-2023-00241934). E.C., J.K.J., C.S., and S.K.Y. are supported by the Korean National Research Foundation (RS-2022-NR070872). J.R. acknowledges support by the Institut de Physique des deux infinis of Sorbonne Université and by the ANR grant ANR-19-CE31-0017 of the French Agence Nationale de la Recherche.
This work was granted access to the HPC resources of KISTI under the allocations KSC-2021-CRE-0486, KSC-2022-CRE-0088, KSC-2022-CRE-0344, KSC-2022-CRE-0409, KSC-2023-CRE-0343, KSC-2024-CHA-0009, and KSC-2025-CRE-0031 and of GENCI under the allocations A0150414625 and A0180416216
The large data transfer was supported by KREONET, which is managed and operated by KISTI.

\section*{Data Availability}
The derived data products used to produce the figures and results presented in this work are available from the corresponding author upon reasonable request. This includes the processed quantities underlying the figures and the galaxy catalog information required to reproduce the analyses presented here. Requests should be directed to the corresponding author and should specify the data products required. The full simulation outputs (\texttt{NewCluster}) are publicly available at \url{https://gemsimulation.com}.


\bibliography{paper}{}

@ARTICLE{han2026,
       author = {{Han}, S. and {Yi}, S.~K. and {Dubois}, Y. and {Rhee}, J. and {Jeon}, S. and {Jang}, J.~K. and {Byun}, G.-H. and {Cadiou}, C. and {Kim}, J. and {Kimm}, T. and {Pichon}, C.},
        title = "{Introducing NewCluster: First half of the history of a high-resolution cluster simulation}",
      journal = {\aap},
         year = 2026,
        month = jan,
       volume = {705},
          eid = {A169},
        pages = {A169},
          doi = {10.1051/0004-6361/202556291},
archivePrefix = {arXiv},
       eprint = {2507.06301},
 primaryClass = {astro-ph.GA},
       adsurl = {https://ui.adsabs.harvard.edu/abs/2026A&A...705A.169H}
}

@ARTICLE{schaye2026,
       author = {{Schaye}, Joop and {Chaikin}, Evgenii and {Schaller}, Matthieu and {Ploeckinger}, Sylvia and {Hu{\v{s}}ko}, Filip and {McGibbon}, Robert J. and {Trayford}, James W. and {Ben{\'\i}tez-Llambay}, Alejandro and {Correa}, Camila and {Frenk}, Carlos S. and {Richings}, Alexander J. and {Forouhar Moreno}, Victor J. and {Bah{\'e}}, Yannick M. and {Borrow}, Josh and {Durrant}, Anna and {Gebek}, Andrea and {Helly}, John C. and {Jenkins}, Adrian and {Lacey}, Cedric G. and {Ludlow}, Aaron and {Nobels}, Folkert S.~J.},
        title = "{The COLIBRE project: cosmological hydrodynamical simulations of galaxy formation and evolution}",
      journal = {\mnras},
         year = 2026,
        month = may,
       volume = {548},
       number = {1},
          eid = {stag375},
        pages = {stag375},
          doi = {10.1093/mnras/stag375},
archivePrefix = {arXiv},
       eprint = {2508.21126},
 primaryClass = {astro-ph.GA},
       adsurl = {https://ui.adsabs.harvard.edu/abs/2026MNRAS.548ag375S}
}

@ARTICLE{jiang2011,
       author = {{Jiang}, Y.-F. and {Greene}, J.~E. and {Ho}, L.~C. and {Xiao}, T. and {Barth}, A.~J.},
        title = "{The Host Galaxies of Low-mass Black Holes}",
      journal = {\apj},
         year = 2011,
        month = dec,
       volume = {742},
          eid = {68},
        pages = {68},
          doi = {10.1088/0004-637X/742/2/68},
archivePrefix = {arXiv},
       eprint = {1107.4105},
 primaryClass = {astro-ph.CO},
       adsurl = {https://ui.adsabs.harvard.edu/abs/2011ApJ...742...68J}
}

@ARTICLE{liu2018,
       author = {{Liu}, H.-Y. and {Yuan}, W. and {Dong}, X.-B. and {Zhou}, H. and {Liu}, W.-J.},
        title = "{A Uniformly Selected Sample of Low-Mass Black Holes in Seyfert 1 Galaxies. II. The SDSS DR7 Sample}",
      journal = {\apjs},
         year = 2018,
        month = apr,
       volume = {235},
          eid = {40},
        pages = {40},
          doi = {10.3847/1538-4365/aab88e},
archivePrefix = {arXiv},
       eprint = {1803.04330},
 primaryClass = {astro-ph.GA},
       adsurl = {https://ui.adsabs.harvard.edu/abs/2018ApJS..235...40L}
}

@ARTICLE{greene2020,
       author = {{Greene}, J.~E. and {Strader}, J. and {Ho}, L.~C.},
        title = "{Intermediate-Mass Black Holes}",
      journal = {\araa},
         year = 2020,
        month = aug,
       volume = {58},
        pages = {257-312},
          doi = {10.1146/annurev-astro-032620-021835},
archivePrefix = {arXiv},
       eprint = {1911.09678},
 primaryClass = {astro-ph.GA},
       adsurl = {https://ui.adsabs.harvard.edu/abs/2020ARA&A..58..257G}
}

@ARTICLE{baldassare2015,
       author = {{Baldassare}, V.~F. and {Reines}, A.~E. and {Gallo}, E. and {Greene}, J.~E.},
        title = "{A $\sim$50,000 $M_{\odot}$ Solar Mass Black Hole in the Nucleus of RGG 118}",
      journal = {\apjl},
         year = 2015,
        month = aug,
       volume = {809},
          eid = {L14},
        pages = {L14},
          doi = {10.1088/2041-8205/809/1/L14},
archivePrefix = {arXiv},
       eprint = {1506.07531},
 primaryClass = {astro-ph.GA},
       adsurl = {https://ui.adsabs.harvard.edu/abs/2015ApJ...809L..14B}
}

@ARTICLE{cisternas2011,
       author = {{Cisternas}, M. and {Jahnke}, K. and {Bongiorno}, A. and {Inskip}, K.~J. and {Impey}, C.~D. and {Koekemoer}, A.~M. and {Merloni}, A. and {Salvato}, M. and {Trump}, J.~R.},
        title = "{Secular Evolution and a Non-evolving Black-hole-to-galaxy Mass Ratio in the Last 7 Gyr}",
      journal = {\apjl},
         year = 2011,
        month = nov,
       volume = {741},
          eid = {L11},
        pages = {L11},
          doi = {10.1088/2041-8205/741/1/L11},
archivePrefix = {arXiv},
       eprint = {1109.4633},
 primaryClass = {astro-ph.CO},
       adsurl = {https://ui.adsabs.harvard.edu/abs/2011ApJ...741L..11C}
}

@ARTICLE{marconi2003,
       author = {{Marconi}, A. and {Hunt}, L.~K.},
        title = "{The Relation between Black Hole Mass, Bulge Mass, and Near-Infrared Luminosity}",
      journal = {\apjl},
         year = 2003,
        month = may,
       volume = {589},
        pages = {L21-L24},
          doi = {10.1086/375804},
archivePrefix = {arXiv},
       eprint = {astro-ph/0304274},
 primaryClass = {astro-ph},
       adsurl = {https://ui.adsabs.harvard.edu/abs/2003ApJ...589L..21M}
}

@ARTICLE{gultekin2009,
       author = {{G{\"u}ltekin}, K. and {Richstone}, D.~O. and {Gebhardt}, K. and {Lauer}, T.~R. and {Tremaine}, S. and {Aller}, M.~C. and {Bender}, R. and {Dressler}, A. and {Faber}, S.~M. and {Filippenko}, A.~V. and {Green}, R. and {Ho}, L.~C. and {Kormendy}, J. and {Magorrian}, J. and {Pinkney}, J. and {Siopis}, C.},
        title = "{The $M$-$\sigma$ and $M$-$L$ Relations in Galactic Bulges, and Determinations of Their Intrinsic Scatter}",
      journal = {\apj},
         year = 2009,
        month = jun,
       volume = {698},
        pages = {198-221},
          doi = {10.1088/0004-637X/698/1/198},
archivePrefix = {arXiv},
       eprint = {0903.4897},
 primaryClass = {astro-ph.GA},
       adsurl = {https://ui.adsabs.harvard.edu/abs/2009ApJ...698..198G}
}

@ARTICLE{sani2011,
       author = {{Sani}, E. and {Marconi}, A. and {Hunt}, L.~K. and {Risaliti}, G.},
        title = "{The Spitzer/IRAC View of Black Hole--Bulge Scaling Relations}",
      journal = {\mnras},
         year = 2011,
        month = may,
       volume = {413},
        pages = {1479-1494},
          doi = {10.1111/j.1365-2966.2011.18229.x},
archivePrefix = {arXiv},
       eprint = {1012.3073},
 primaryClass = {astro-ph.CO},
       adsurl = {https://ui.adsabs.harvard.edu/abs/2011MNRAS.413.1479S}
}

@ARTICLE{lasker2014,
       author = {{L{\"a}sker}, R. and {Ferrarese}, L. and {van de Ven}, G. and {Shankar}, F.},
        title = "{Supermassive Black Holes and Their Host Galaxies. II. The Correlation with Near-infrared Luminosity Revisited}",
      journal = {\apj},
         year = 2014,
        month = jan,
       volume = {780},
          eid = {70},
        pages = {70},
          doi = {10.1088/0004-637X/780/1/70},
archivePrefix = {arXiv},
       eprint = {1311.1531},
 primaryClass = {astro-ph.GA},
       adsurl = {https://ui.adsabs.harvard.edu/abs/2014ApJ...780...70L}
}

@ARTICLE{savorgnan2016,
       author = {{Savorgnan}, G.~A.~D. and {Graham}, A.~W. and {Marconi}, A. and {Sani}, E.},
        title = "{Supermassive Black Holes and Their Host Spheroids. II. The Red and Blue Sequence in the $M_{\rm BH}$--$M_{*,\rm sph}$ Diagram}",
      journal = {\apj},
         year = 2016,
        month = jan,
       volume = {817},
          eid = {21},
        pages = {21},
          doi = {10.3847/0004-637X/817/1/21},
archivePrefix = {arXiv},
       eprint = {1511.07437},
 primaryClass = {astro-ph.GA},
       adsurl = {https://ui.adsabs.harvard.edu/abs/2016ApJ...817...21S}
}

@ARTICLE{kormendy2011,
       author = {{Kormendy}, J. and {Bender}, R. and {Cornell}, M.~E.},
        title = "{Supermassive Black Holes Do Not Correlate with Galaxy Disks or Pseudobulges}",
      journal = {\nat},
         year = 2011,
        month = jan,
       volume = {469},
        pages = {374-376},
          doi = {10.1038/nature09694},
archivePrefix = {arXiv},
       eprint = {1101.3781},
 primaryClass = {astro-ph.CO},
       adsurl = {https://ui.adsabs.harvard.edu/abs/2011Natur.469..374K}
}

@ARTICLE{graham2013,
       author = {{Graham}, A.~W. and {Scott}, N.},
        title = "{The (Black Hole Mass)--(Host Spheroid Luminosity) Relation at High and Low Masses, the Quadratic Growth of Black Holes, and Intermediate-mass Black Hole Candidates}",
      journal = {\apj},
         year = 2013,
        month = feb,
       volume = {764},
          eid = {151},
        pages = {151},
          doi = {10.1088/0004-637X/764/2/151},
archivePrefix = {arXiv},
       eprint = {1211.3199},
 primaryClass = {astro-ph.CO},
       adsurl = {https://ui.adsabs.harvard.edu/abs/2013ApJ...764..151G}
}

@ARTICLE{scott2013,
       author = {{Scott}, N. and {Graham}, A.~W. and {Schombert}, J.},
        title = "{The Supermassive Black Hole Mass--Spheroid Stellar Mass Relation for S{\'e}rsic and Core-S{\'e}rsic Galaxies}",
      journal = {\apj},
         year = 2013,
        month = may,
       volume = {768},
          eid = {76},
        pages = {76},
          doi = {10.1088/0004-637X/768/1/76},
archivePrefix = {arXiv},
       eprint = {1303.5490},
 primaryClass = {astro-ph.CO},
       adsurl = {https://ui.adsabs.harvard.edu/abs/2013ApJ...768...76S}
}

@ARTICLE{sahu2019,
       author = {{Sahu}, N. and {Graham}, A.~W. and {Davis}, B.~L.},
        title = "{Black Hole Mass Scaling Relations for Early-type Galaxies. I. $M_{\rm BH}$--$M_{*,\rm sph}$ and $M_{\rm BH}$--$M_{*,\rm gal}$}",
      journal = {\apj},
         year = 2019,
        month = may,
       volume = {876},
          eid = {155},
        pages = {155},
          doi = {10.3847/1538-4357/ab0f32},
archivePrefix = {arXiv},
       eprint = {1903.04738},
 primaryClass = {astro-ph.GA},
       adsurl = {https://ui.adsabs.harvard.edu/abs/2019ApJ...876..155S}
}

@ARTICLE{lauer2007,
       author = {{Lauer}, T.~R. and {Tremaine}, S. and {Richstone}, D. and {Faber}, S.~M.},
        title = "{Selection Bias in Observing the Cosmological Evolution of the $M_{\bullet}$-$\sigma$ and $M_{\bullet}$-$L$ Relationships}",
      journal = {\apj},
         year = 2007,
        month = nov,
       volume = {670},
        pages = {249-260},
          doi = {10.1086/522083},
archivePrefix = {arXiv},
       eprint = {0705.4103},
 primaryClass = {astro-ph},
       adsurl = {https://ui.adsabs.harvard.edu/abs/2007ApJ...670..249L}
}

@ARTICLE{schulze2011,
       author = {{Schulze}, A. and {Wisotzki}, L.},
        title = "{Selection Effects in the Black Hole--Bulge Relation and Its Evolution}",
      journal = {\aap},
         year = 2011,
        month = nov,
       volume = {535},
          eid = {A87},
        pages = {A87},
          doi = {10.1051/0004-6361/201117564},
archivePrefix = {arXiv},
       eprint = {1109.2787},
 primaryClass = {astro-ph.CO},
       adsurl = {https://ui.adsabs.harvard.edu/abs/2011A&A...535A..87S}
}

@ARTICLE{schulze2014,
       author = {{Schulze}, A. and {Wisotzki}, L.},
        title = "{Accounting for Selection Effects in the BH--Bulge Relations: No Evidence for Cosmological Evolution}",
      journal = {\mnras},
         year = 2014,
        month = mar,
       volume = {438},
        pages = {3422-3433},
          doi = {10.1093/mnras/stt2457},
archivePrefix = {arXiv},
       eprint = {1312.5610},
 primaryClass = {astro-ph.CO},
       adsurl = {https://ui.adsabs.harvard.edu/abs/2014MNRAS.438.3422S}
}

@ARTICLE{peng2006,
       author = {{Peng}, C.~Y. and {Impey}, C.~D. and {Rix}, H.-W. and {Kochanek}, C.~S. and {Keeton}, C.~R. and {Falco}, E.~E. and {Leh{\'a}r}, J. and {McLeod}, B.~A.},
        title = "{Probing the Coevolution of Supermassive Black Holes and Galaxies Using Gravitationally Lensed Quasar Hosts}",
      journal = {\apj},
         year = 2006,
        month = oct,
       volume = {649},
        pages = {616-634},
          doi = {10.1086/506266},
archivePrefix = {arXiv},
       eprint = {astro-ph/0603248},
 primaryClass = {astro-ph},
       adsurl = {https://ui.adsabs.harvard.edu/abs/2006ApJ...649..616P}
}

@ARTICLE{jahnke2009,
       author = {{Jahnke}, K. and {Bongiorno}, A. and {Brusa}, M. and {Capak}, P. and {Cappelluti}, N. and {Cisternas}, M. and {Civano}, F. and {Colbert}, J. and {Comastri}, A. and {Elvis}, M. and {Hasinger}, G. and {Ilbert}, O. and {Impey}, C. and {Inskip}, K. and {Koekemoer}, A.~M. and {Lilly}, S. and {Maier}, C. and {Merloni}, A. and {Riechers}, D. and {Salvato}, M. and {Schinnerer}, E. and {Scoville}, N.~Z. and {Silverman}, J. and {Taniguchi}, Y. and {Trump}, J.~R. and {Yan}, L.},
        title = "{Massive Galaxies in COSMOS: Evolution of Black Hole versus Bulge Mass but Not versus Total Stellar Mass over the Last 9 Gyr?}",
      journal = {\apjl},
         year = 2009,
        month = nov,
       volume = {706},
        pages = {L215-L220},
          doi = {10.1088/0004-637X/706/2/L215},
archivePrefix = {arXiv},
       eprint = {0907.5199},
 primaryClass = {astro-ph.CO},
       adsurl = {https://ui.adsabs.harvard.edu/abs/2009ApJ...706L.215J}
}

@ARTICLE{suh2020,
       author = {{Suh}, H. and {Civano}, F. and {Trakhtenbrot}, B. and {Shankar}, F. and {Hasinger}, G. and {Sanders}, D.~B. and {Allevato}, V.},
        title = "{No Significant Evolution of Relations between Black Hole Mass and Galaxy Total Stellar Mass Up to $z\sim2.5$}",
      journal = {\apj},
         year = 2020,
        month = jan,
       volume = {889},
          eid = {32},
        pages = {32},
          doi = {10.3847/1538-4357/ab5f5f},
archivePrefix = {arXiv},
       eprint = {1912.02824},
 primaryClass = {astro-ph.GA},
       adsurl = {https://ui.adsabs.harvard.edu/abs/2020ApJ...889...32S}
}

@ARTICLE{ding2022,
       author = {{Ding}, X. and {Silverman}, J.~D. and {Onoue}, M.},
        title = "{Opening the Era of Quasar-host Studies at High Redshift with JWST}",
      journal = {\apjl},
         year = 2022,
        month = nov,
       volume = {939},
          eid = {L28},
        pages = {L28},
          doi = {10.3847/2041-8213/ac9c02},
archivePrefix = {arXiv},
       eprint = {2209.03359},
 primaryClass = {astro-ph.GA},
       adsurl = {https://ui.adsabs.harvard.edu/abs/2022ApJ...939L..28D}
}

@ARTICLE{dimatteo2008,
       author = {{Di Matteo}, T. and {Colberg}, J. and {Springel}, V. and {Hernquist}, L. and {Sijacki}, D.},
        title = "{Direct Cosmological Simulations of the Growth of Black Holes and Galaxies}",
      journal = {\apj},
         year = 2008,
        month = mar,
       volume = {676},
        pages = {33-53},
          doi = {10.1086/524921},
archivePrefix = {arXiv},
       eprint = {0705.2269},
 primaryClass = {astro-ph},
       adsurl = {https://ui.adsabs.harvard.edu/abs/2008ApJ...676...33D}
}

@ARTICLE{khandai2015,
       author = {{Khandai}, N. and {Di Matteo}, T. and {Croft}, R. and {Wilkins}, S.~M. and {Feng}, Y. and {Tucker}, E. and {DeGraf}, C. and {Liu}, M.-S.},
        title = "{The MassiveBlack-II Simulation: The Evolution of Haloes and Galaxies to $z\sim0$}",
      journal = {\mnras},
         year = 2015,
        month = jun,
       volume = {450},
        pages = {1349-1374},
          doi = {10.1093/mnras/stv627},
archivePrefix = {arXiv},
       eprint = {1402.0888},
 primaryClass = {astro-ph.CO},
       adsurl = {https://ui.adsabs.harvard.edu/abs/2015MNRAS.450.1349K}
}

@ARTICLE{sijacki2015,
       author = {{Sijacki}, D. and {Vogelsberger}, M. and {Genel}, S. and {Springel}, V. and {Torrey}, P. and {Snyder}, G.~F. and {Nelson}, D. and {Hernquist}, L.},
        title = "{The Illustris Simulation: The Evolving Population of Black Holes across Cosmic Time}",
      journal = {\mnras},
         year = 2015,
        month = sep,
       volume = {452},
        pages = {575-596},
          doi = {10.1093/mnras/stv1340},
archivePrefix = {arXiv},
       eprint = {1408.6842},
 primaryClass = {astro-ph.GA},
       adsurl = {https://ui.adsabs.harvard.edu/abs/2015MNRAS.452..575S}
}

@ARTICLE{dubois2016,
       author = {{Dubois}, Y. and {Peirani}, S. and {Pichon}, C. and {Devriendt}, J. and {Gavazzi}, R. and {Welker}, C. and {Volonteri}, M.},
        title = "{The Horizon-AGN Simulation: Morphological Diversity of Galaxies Promoted by AGN Feedback}",
      journal = {\mnras},
         year = 2016,
        month = dec,
       volume = {463},
        pages = {3948-3964},
          doi = {10.1093/mnras/stw2265},
archivePrefix = {arXiv},
       eprint = {1606.03086},
 primaryClass = {astro-ph.GA},
       adsurl = {https://ui.adsabs.harvard.edu/abs/2016MNRAS.463.3948D}
}

@ARTICLE{weinberger2018,
       author = {{Weinberger}, R. and {Springel}, V. and {Pakmor}, R. and {Nelson}, D. and {Genel}, S. and {Pillepich}, A. and {Vogelsberger}, M. and {Marinacci}, F. and {Naiman}, J. and {Torrey}, P. and {Hernquist}, L.},
        title = "{Supermassive Black Holes and Their Feedback Effects in the IllustrisTNG Simulation}",
      journal = {\mnras},
         year = 2018,
        month = sep,
       volume = {479},
        pages = {4056-4072},
          doi = {10.1093/mnras/sty1733},
archivePrefix = {arXiv},
       eprint = {1710.04659},
 primaryClass = {astro-ph.GA},
       adsurl = {https://ui.adsabs.harvard.edu/abs/2018MNRAS.479.4056W}
}

@ARTICLE{dave2019,
       author = {{Dav{\'e}}, R. and {Angl{\'e}s-Alc{\'a}zar}, D. and {Narayanan}, D. and {Li}, Q. and {Rafieferantsoa}, M.~H. and {Appleby}, S.},
        title = "{SIMBA: Cosmological Simulations with Black Hole Growth and Feedback}",
      journal = {\mnras},
         year = 2019,
        month = jun,
       volume = {486},
        pages = {2827-2849},
          doi = {10.1093/mnras/stz937},
archivePrefix = {arXiv},
       eprint = {1901.10203},
 primaryClass = {astro-ph.GA},
       adsurl = {https://ui.adsabs.harvard.edu/abs/2019MNRAS.486.2827D}
}

@ARTICLE{gallo2010,
       author = {{Gallo}, E. and {Treu}, T. and {Marshall}, P.~J. and {Woo}, J.-H. and {Leipski}, C. and {Antonucci}, R.},
        title = "{AMUSE-Virgo. II. Down-sizing in Black Hole Accretion}",
      journal = {\apj},
         year = 2010,
        month = may,
       volume = {714},
        pages = {25-36},
          doi = {10.1088/0004-637X/714/1/25},
archivePrefix = {arXiv},
       eprint = {1002.3619},
 primaryClass = {astro-ph.CO},
       adsurl = {https://ui.adsabs.harvard.edu/abs/2010ApJ...714...25G}
}

@ARTICLE{reines2013,
       author = {{Reines}, A.~E. and {Greene}, J.~E. and {Geha}, M.},
        title = "{Dwarf Galaxies with Optical Signatures of Active Massive Black Holes}",
      journal = {\apj},
         year = 2013,
        month = oct,
       volume = {775},
          eid = {116},
        pages = {116},
          doi = {10.1088/0004-637X/775/2/116},
archivePrefix = {arXiv},
       eprint = {1308.0328},
 primaryClass = {astro-ph.CO},
       adsurl = {https://ui.adsabs.harvard.edu/abs/2013ApJ...775..116R}
}

@ARTICLE{mezcua2018,
       author = {{Mezcua}, M. and {Civano}, F. and {Marchesi}, S. and {Suh}, H. and {Fabbiano}, G. and {Volonteri}, M.},
        title = "{Intermediate-mass Black Holes in Dwarf Galaxies out to Redshift $\sim2.4$ in the Chandra COSMOS-Legacy Survey}",
      journal = {\mnras},
         year = 2018,
        month = aug,
       volume = {478},
        pages = {2576-2591},
          doi = {10.1093/mnras/sty1163},
archivePrefix = {arXiv},
       eprint = {1802.01567},
 primaryClass = {astro-ph.GA},
       adsurl = {https://ui.adsabs.harvard.edu/abs/2018MNRAS.478.2576M}
}

@ARTICLE{tanaka2024,
       author = {{Tanaka}, T.~S. and {Silverman}, J.~D. and {Ding}, X. and {Jahnke}, K. and {Trakhtenbrot}, B. and {Lambrides}, E. and {Onoue}, M. and {Andika}, I.~T. and {Bongiorno}, A. and {Faisst}, A.~L. and {Gillman}, S. and {Hayward}, C.~C. and {Hirschmann}, M. and {Koekemoer}, A. and {Kokorev}, V. and {Liu}, Z. and {Magdis}, G.~E. and {Renzini}, A. and {Casey}, C. and {Drakos}, N.~E. and {Franco}, M. and {Gozaliasl}, G. and {Kartaltepe}, J. and {Liu}, D. and {McCracken}, H.~J. and {Rhodes}, J. and {Robertson}, B. and {Toft}, S.},
        title = "{The $M_{\rm BH}$--$M_{\star}$ Relation up to $z\sim2$ through Decomposition of COSMOS-Web NIRCam Images}",
      journal = {\apj},
         year = 2025,
        month = feb,
       volume = {979},
          eid = {215},
        pages = {215},
          doi = {10.3847/1538-4357/ad9d0a},
archivePrefix = {arXiv},
       eprint = {2401.13742},
 primaryClass = {astro-ph.GA},
       adsurl = {https://ui.adsabs.harvard.edu/abs/2025ApJ...979..215T}
}

@ARTICLE{li2024,
       author = {{Li}, J. and {Silverman}, J.~D. and {Shen}, Y. and {Volonteri}, M. and {Jahnke}, K. and {Zhuang}, M.-Y. and {Scoggins}, M.~T. and {Ding}, X. and {Harikane}, Y. and {Onoue}, M. and {Tanaka}, T.~S.},
        title = "{Tip of the Iceberg: Overmassive Black Holes at $4<z<7$ Found by JWST Are Not Inconsistent with the Local $M_{\rm BH}$--$M_{\star}$ Relation}",
      journal = {\apj},
         year = 2025,
        month = mar,
       volume = {981},
          eid = {19},
        pages = {19},
          doi = {10.3847/1538-4357/ada603},
archivePrefix = {arXiv},
       eprint = {2403.00074},
 primaryClass = {astro-ph.GA},
       adsurl = {https://ui.adsabs.harvard.edu/abs/2025ApJ...981...19L}
}

@ARTICLE{magorrian1998,
       author = {{Magorrian}, J. and {Tremaine}, S. and {Richstone}, D. and {Bender}, R. and {Bower}, G. and {Dressler}, A. and {Faber}, S.~M. and {Gebhardt}, K. and {Green}, R. and {Grillmair}, C. and {Kormendy}, J. and {Lauer}, T.},
        title = "{The Demography of Massive Dark Objects in Galaxy Centers}",
      journal = {\aj},
         year = 1998,
        month = jun,
       volume = {115},
        pages = {2285-2305},
          doi = {10.1086/300353},
archivePrefix = {arXiv},
       eprint = {astro-ph/9708072},
 primaryClass = {astro-ph},
       adsurl = {https://ui.adsabs.harvard.edu/abs/1998AJ....115.2285M}
}

@ARTICLE{ferrarese2000,
       author = {{Ferrarese}, L. and {Merritt}, D.},
        title = "{A Fundamental Relation between Supermassive Black Holes and Their Host Galaxies}",
      journal = {\apjl},
         year = 2000,
        month = aug,
       volume = {539},
        pages = {L9-L12},
          doi = {10.1086/312838},
archivePrefix = {arXiv},
       eprint = {astro-ph/0006053},
 primaryClass = {astro-ph},
       adsurl = {https://ui.adsabs.harvard.edu/abs/2000ApJ...539L...9F}
}

@ARTICLE{gebhardt2000,
       author = {{Gebhardt}, K. and {Bender}, R. and {Bower}, G. and {Dressler}, A. and {Faber}, S.~M. and {Filippenko}, A.~V. and {Green}, R. and {Grillmair}, C. and {Ho}, L.~C. and {Kormendy}, J. and {Lauer}, T.~R. and {Magorrian}, J. and {Pinkney}, J. and {Richstone}, D. and {Tremaine}, S.},
        title = "{A Relationship between Nuclear Black Hole Mass and Galaxy Velocity Dispersion}",
      journal = {\apjl},
         year = 2000,
        month = aug,
       volume = {539},
        pages = {L13-L16},
          doi = {10.1086/312840},
archivePrefix = {arXiv},
       eprint = {astro-ph/0006289},
 primaryClass = {astro-ph},
       adsurl = {https://ui.adsabs.harvard.edu/abs/2000ApJ...539L..13G}
}

@ARTICLE{tremaine2002,
       author = {{Tremaine}, S. and {Gebhardt}, K. and {Bender}, R. and {Bower}, G. and {Dressler}, A. and {Faber}, S.~M. and {Filippenko}, A.~V. and {Green}, R. and {Grillmair}, C. and {Ho}, L.~C. and {Kormendy}, J. and {Lauer}, T.~R. and {Magorrian}, J. and {Pinkney}, J. and {Richstone}, D.},
        title = "{The Slope of the Black Hole Mass versus Velocity Dispersion Correlation}",
      journal = {\apj},
         year = 2002,
        month = aug,
       volume = {574},
        pages = {740-753},
          doi = {10.1086/341002},
archivePrefix = {arXiv},
       eprint = {astro-ph/0203468},
 primaryClass = {astro-ph},
       adsurl = {https://ui.adsabs.harvard.edu/abs/2002ApJ...574..740T}
}

@ARTICLE{haring2004,
       author = {{H{\"a}ring}, N. and {Rix}, H.-W.},
        title = "{On the Black Hole Mass--Bulge Mass Relation}",
      journal = {\apjl},
         year = 2004,
        month = apr,
       volume = {604},
        pages = {L89-L92},
          doi = {10.1086/383567},
archivePrefix = {arXiv},
       eprint = {astro-ph/0402376},
 primaryClass = {astro-ph},
       adsurl = {https://ui.adsabs.harvard.edu/abs/2004ApJ...604L..89H}
}

@ARTICLE{mcconnell2013,
       author = {{McConnell}, N.~J. and {Ma}, C.-P.},
        title = "{Revisiting the Scaling Relations of Black Hole Masses and Host Galaxy Properties}",
      journal = {\apj},
         year = 2013,
        month = feb,
       volume = {764},
          eid = {184},
        pages = {184},
          doi = {10.1088/0004-637X/764/2/184},
archivePrefix = {arXiv},
       eprint = {1211.2816},
 primaryClass = {astro-ph.CO},
       adsurl = {https://ui.adsabs.harvard.edu/abs/2013ApJ...764..184M}
}

@ARTICLE{kormendy2013,
       author = {{Kormendy}, J. and {Ho}, L.~C.},
        title = "{Coevolution (Or Not) of Supermassive Black Holes and Host Galaxies}",
      journal = {\araa},
         year = 2013,
       volume = {51},
        pages = {511-653},
          doi = {10.1146/annurev-astro-082708-101811},
archivePrefix = {arXiv},
       eprint = {1304.7762},
 primaryClass = {astro-ph.CO},
       adsurl = {https://ui.adsabs.harvard.edu/abs/2013ARA&A..51..511K}
}

@ARTICLE{bentz2018,
       author = {{Bentz}, M.~C. and {Manne-Nicholas}, E.},
        title = "{Black Hole--Galaxy Scaling Relationships for Active Galactic Nuclei with Reverberation Masses}",
      journal = {\apj},
         year = 2018,
       volume = {864},
          eid = {146},
        pages = {146},
          doi = {10.3847/1538-4357/aad808},
archivePrefix = {arXiv},
       eprint = {1808.01329},
 primaryClass = {astro-ph.GA},
       adsurl = {https://ui.adsabs.harvard.edu/abs/2018ApJ...864..146B}
}

@ARTICLE{sturm2024,
       author = {{Sturm}, M.~R. and {Reines}, A.~E.},
        title = "{A Breakdown of the Black Hole--Bulge Mass Relation in Local Active Galaxies}",
      journal = {\apj},
         year = 2024,
        month = aug,
       volume = {971},
          eid = {173},
        pages = {173},
          doi = {10.3847/1538-4357/ad55e6},
archivePrefix = {arXiv},
       eprint = {2406.06675},
 primaryClass = {astro-ph.GA},
       adsurl = {https://ui.adsabs.harvard.edu/abs/2024ApJ...971..173S}
}

@ARTICLE{davis2018,
       author = {{Davis}, B.~L. and {Graham}, A.~W. and {Cameron}, E.},
        title = "{Black Hole Mass Scaling Relations for Spiral Galaxies. II. $M_{\rm BH}$--$M_{\rm *,tot}$ and $M_{\rm BH}$--$M_{\rm *,disk}$}",
      journal = {\apj},
         year = 2018,
        month = dec,
       volume = {869},
          eid = {113},
        pages = {113},
          doi = {10.3847/1538-4357/aae820},
       adsurl = {https://ui.adsabs.harvard.edu/abs/2018ApJ...869..113D}
}

@ARTICLE{davis2019,
       author = {{Davis}, B.~L. and {Graham}, A.~W. and {Cameron}, E.},
        title = "{Black Hole Mass Scaling Relations for Spiral Galaxies. I. $M_{\rm BH}$--$M_{\rm *,sph}$}",
      journal = {\apj},
         year = 2019,
        month = mar,
       volume = {873},
          eid = {85},
        pages = {85},
          doi = {10.3847/1538-4357/aaf3b8},
       adsurl = {https://ui.adsabs.harvard.edu/abs/2019ApJ...873...85D}
}

@ARTICLE{shankar2016,
       author = {{Shankar}, F. and {Bernardi}, M. and {Sheth}, R.~K. and {Ferrarese}, L. and {Graham}, A.~W. and {Savorgnan}, G. and {Allevato}, V. and {Marconi}, A. and {L{\"a}sker}, R. and {Lapi}, A.},
        title = "{Selection Bias in Dynamically Measured Supermassive Black Hole Samples: Its Consequences and the Quest for the Most Fundamental Relation}",
      journal = {\mnras},
         year = 2016,
        month = aug,
       volume = {460},
        pages = {3119-3142},
          doi = {10.1093/mnras/stw678},
archivePrefix = {arXiv},
       eprint = {1603.01276},
 primaryClass = {astro-ph.GA},
       adsurl = {https://ui.adsabs.harvard.edu/abs/2016MNRAS.460.3119S}
}

@ARTICLE{shankar2017,
       author = {{Shankar}, F. and {Bernardi}, M. and {Sheth}, R.~K.},
        title = "{Selection Bias in Dynamically Measured Supermassive Black Hole Samples: Dynamical Masses and Dependence on S{\'e}rsic Index}",
      journal = {\mnras},
         year = 2017,
        month = may,
       volume = {466},
        pages = {4029-4039},
          doi = {10.1093/mnras/stw3368},
archivePrefix = {arXiv},
       eprint = {1701.01732},
 primaryClass = {astro-ph.GA},
       adsurl = {https://ui.adsabs.harvard.edu/abs/2017MNRAS.466.4029S}
}

@ARTICLE{shankar2019,
       author = {{Shankar}, F. and {Bernardi}, M. and {Richardson}, K. and {Marsden}, C. and {Sheth}, R.~K. and {Allevato}, V. and {Graziani}, L. and {Mezcua}, M. and {Ricci}, F. and {Penny}, S.~J. and {La Franca}, F. and {Pacucci}, F.},
        title = "{Black Hole Scaling Relations of Active and Quiescent Galaxies: Addressing Selection Effects and Constraining Virial Factors}",
      journal = {\mnras},
         year = 2019,
       volume = {485},
        pages = {1278-1292},
          doi = {10.1093/mnras/stz376},
archivePrefix = {arXiv},
       eprint = {1901.11036},
 primaryClass = {astro-ph.GA},
       adsurl = {https://ui.adsabs.harvard.edu/abs/2019MNRAS.485.1278S}
}

@ARTICLE{degraf2015,
       author = {{DeGraf}, C. and {Di Matteo}, T. and {Treu}, T. and {Feng}, Y. and {Woo}, J.-H. and {Park}, D.},
        title = "{Scaling Relations between Black Holes and Their Host Galaxies: Comparing Theoretical and Observational Measurements, and the Impact of Selection Effects}",
      journal = {\mnras},
         year = 2015,
        month = nov,
       volume = {454},
        pages = {913-932},
          doi = {10.1093/mnras/stv2002},
archivePrefix = {arXiv},
       eprint = {1412.4133},
 primaryClass = {astro-ph.GA},
       adsurl = {https://ui.adsabs.harvard.edu/abs/2015MNRAS.454..913D}
}

@ARTICLE{merloni2010,
       author = {{Merloni}, A. and {Bongiorno}, A. and {Bolzonella}, M. and {Brusa}, M. and {Civano}, F. and {Comastri}, A. and {Elvis}, M. and {Fiore}, F. and {Gilli}, R. and {Hao}, H. and {Jahnke}, K. and {Koekemoer}, A.~M. and {Lusso}, E. and {Mainieri}, V. and {Mignoli}, M. and {Miyaji}, T. and {Renzini}, A. and {Salvato}, M. and {Silverman}, J. and {Trump}, J. and {Vignali}, C. and {Zamorani}, G. and {Capak}, P. and {Lilly}, S.~J. and {Sanders}, D. and {Taniguchi}, Y. and {Bardelli}, S. and {Carollo}, C.~M. and {Caputi}, K. and {Contini}, T. and {Coppa}, G. and {Cucciati}, O. and {de la Torre}, S. and {de Ravel}, L. and {Franzetti}, P. and {Garilli}, B. and {Hasinger}, G. and {Impey}, C. and {Iovino}, A. and {Iwasawa}, K. and {Kampczyk}, P. and {Kneib}, J.-P. and {Knobel}, C. and {Kova{\v c}}, K. and {Lamareille}, F. and {Le Borgne}, J.-F. and {Le Brun}, V. and {Le F{\`e}vre}, O. and {Maier}, C. and {Pell{\'o}}, R. and {Peng}, Y. and {P{\'e}rez-Montero}, E. and {Ricciardelli}, E. and {Scodeggio}, M. and {Tanaka}, M. and {Tasca}, L.~A.~M. and {Tresse}, L. and {Vergani}, D. and {Zucca}, E.},
        title = "{On the Cosmic Evolution of the Scaling Relations between Black Holes and Their Host Galaxies: Broad-line Active Galactic Nuclei in the zCOSMOS Survey}",
      journal = {\apj},
         year = 2010,
        month = jan,
       volume = {708},
        pages = {137-157},
          doi = {10.1088/0004-637X/708/1/137},
archivePrefix = {arXiv},
       eprint = {0910.4970},
 primaryClass = {astro-ph.CO},
       adsurl = {https://ui.adsabs.harvard.edu/abs/2010ApJ...708..137M}
}

@ARTICLE{bennert2011,
       author = {{Bennert}, V.~N. and {Auger}, M.~W. and {Treu}, T. and {Woo}, J.-H. and {Malkan}, M.~A.},
        title = "{The Relation between Black Hole Mass and Host Spheroid Stellar Mass out to $z\sim2$}",
      journal = {\apj},
         year = 2011,
        month = dec,
       volume = {742},
          eid = {107},
        pages = {107},
          doi = {10.1088/0004-637X/742/2/107},
archivePrefix = {arXiv},
       eprint = {1102.1975},
 primaryClass = {astro-ph.CO},
       adsurl = {https://ui.adsabs.harvard.edu/abs/2011ApJ...742..107B}
}

@ARTICLE{ding2020,
       author = {{Ding}, X. and {Silverman}, J.~D. and {Treu}, T. and {Schulze}, A. and {Schramm}, M. and {Birrer}, S. and {Park}, D. and {Jahnke}, K. and {Bennert}, V.~N. and {Kartaltepe}, J.~S. and {Koekemoer}, A.~M. and {Malkan}, M.~A. and {Sanders}, D.},
        title = "{The Mass Relations between Supermassive Black Holes and Their Host Galaxies at $1<z<2$ with HST-WFC3}",
      journal = {\apj},
         year = 2020,
        month = jan,
       volume = {888},
          eid = {37},
        pages = {37},
          doi = {10.3847/1538-4357/ab5b90},
archivePrefix = {arXiv},
       eprint = {1910.11875},
 primaryClass = {astro-ph.GA},
       adsurl = {https://ui.adsabs.harvard.edu/abs/2020ApJ...888...37D}
}

@ARTICLE{pacucci2023,
       author = {{Pacucci}, F. and {Nguyen}, B. and {Carniani}, S. and {Maiolino}, R. and {Fan}, X.},
        title = "{JWST CEERS and JADES Active Galaxies at $z=4$--7 Violate the Local $M_{\bullet}$--$M_{\star}$ Relation at $>3\sigma$: Implications for Low-mass Black Holes and Seeding Models}",
      journal = {\apjl},
         year = 2023,
        month = oct,
       volume = {957},
          eid = {L3},
        pages = {L3},
          doi = {10.3847/2041-8213/ad0158},
archivePrefix = {arXiv},
       eprint = {2308.12331},
 primaryClass = {astro-ph.GA},
       adsurl = {https://ui.adsabs.harvard.edu/abs/2023ApJ...957L...3P}
}

@ARTICLE{mezcua2024,
       author = {{Mezcua}, M. and {Pacucci}, F. and {Suh}, H. and {Siudek}, M. and {Natarajan}, P.},
        title = "{Overmassive Black Holes at Cosmic Noon: Linking the Local and the High-redshift Universe}",
      journal = {\apjl},
         year = 2024,
       volume = {966},
          eid = {L30},
        pages = {L30},
          doi = {10.3847/2041-8213/ad3c2a},
archivePrefix = {arXiv},
       eprint = {2404.05793},
 primaryClass = {astro-ph.GA},
       adsurl = {https://ui.adsabs.harvard.edu/abs/2024ApJ...966L..30M}
}

@ARTICLE{rosasguevara2016,
       author = {{Rosas-Guevara}, Y. and {Bower}, R.~G. and {Schaye}, J. and {McAlpine}, S. and {Dalla Vecchia}, C. and {Frenk}, C.~S. and {Schaller}, M. and {Theuns}, T.},
        title = "{Supermassive Black Holes in the EAGLE Universe. Revealing the Observables of Their Growth}",
      journal = {\mnras},
         year = 2016,
        month = oct,
       volume = {462},
        pages = {190-205},
          doi = {10.1093/mnras/stw1679},
archivePrefix = {arXiv},
       eprint = {1604.00020},
 primaryClass = {astro-ph.GA},
       adsurl = {https://ui.adsabs.harvard.edu/abs/2016MNRAS.462..190R}
}

@ARTICLE{thomas2019,
       author = {{Thomas}, N. and {Dav{\'e}}, R. and {Angl{\'e}s-Alc{\'a}zar}, D. and {Jarvis}, M.},
        title = "{Black Hole--Galaxy Correlations in SIMBA}",
      journal = {\mnras},
         year = 2019,
        month = aug,
       volume = {487},
        pages = {5764-5780},
          doi = {10.1093/mnras/stz1703},
archivePrefix = {arXiv},
       eprint = {1905.02741},
 primaryClass = {astro-ph.GA},
       adsurl = {https://ui.adsabs.harvard.edu/abs/2019MNRAS.487.5764T}
}

@ARTICLE{li2020,
       author = {{Li}, Y. and {Habouzit}, M. and {Genel}, S. and {Somerville}, R.~S. and {Terrazas}, B.~A. and {Bell}, E.~F. and {Pillepich}, A. and {Nelson}, D. and {Weinberger}, R. and {Rodriguez-Gomez}, V. and {Ma}, C.-P. and {Pakmor}, R. and {Hernquist}, L. and {Vogelsberger}, M.},
        title = "{Correlations between Black Holes and Host Galaxies in the Illustris and IllustrisTNG Simulations}",
      journal = {\apj},
         year = 2020,
       volume = {895},
          eid = {102},
        pages = {102},
          doi = {10.3847/1538-4357/ab8f8d},
archivePrefix = {arXiv},
       eprint = {1910.00017},
 primaryClass = {astro-ph.GA},
       adsurl = {https://ui.adsabs.harvard.edu/abs/2020ApJ...895..102L}
}

@ARTICLE{habouzit2021,
       author = {{Habouzit}, M. and {Li}, Y. and {Somerville}, R.~S. and {Genel}, S. and {Pillepich}, A. and {Volonteri}, M. and {Dav{\'e}}, R. and {Rosas-Guevara}, Y. and {McAlpine}, S. and {Peirani}, S. and {Hernquist}, L. and {Angl{\'e}s-Alc{\'a}zar}, D. and {Reines}, A. and {Bower}, R. and {Dubois}, Y. and {Nelson}, D. and {Pichon}, C. and {Vogelsberger}, M.},
        title = "{Supermassive Black Holes in Cosmological Simulations I: $M_{\rm BH}$--$M_{\star}$ Relation and Black Hole Mass Function}",
      journal = {\mnras},
         year = 2021,
        month = may,
       volume = {503},
        pages = {1940-1975},
          doi = {10.1093/mnras/stab496},
archivePrefix = {arXiv},
       eprint = {2006.10094},
 primaryClass = {astro-ph.GA},
       adsurl = {https://ui.adsabs.harvard.edu/abs/2021MNRAS.503.1940H}
}

@ARTICLE{terrazas2020,
       author = {{Terrazas}, B.~A. and {Bell}, E.~F. and {Pillepich}, A. and {Nelson}, D. and {Somerville}, R.~S. and {Genel}, S. and {Weinberger}, R. and {Habouzit}, M. and {Li}, Y. and {Hernquist}, L. and {Vogelsberger}, M.},
        title = "{The Relationship between Black Hole Mass and Galaxy Properties: Examining the Black Hole Feedback Model in IllustrisTNG}",
      journal = {\mnras},
         year = 2020,
       volume = {493},
        pages = {1888-1906},
          doi = {10.1093/mnras/staa374},
archivePrefix = {arXiv},
       eprint = {1906.02747},
 primaryClass = {astro-ph.GA},
       adsurl = {https://ui.adsabs.harvard.edu/abs/2020MNRAS.493.1888T}
}

@ARTICLE{somerville2008,
       author = {{Somerville}, R.~S. and {Hopkins}, P.~F. and {Cox}, T.~J. and {Robertson}, B.~E. and {Hernquist}, L.},
        title = "{A Semi-analytic Model for the Co-evolution of Galaxies, Black Holes and Active Galactic Nuclei}",
      journal = {\mnras},
         year = 2008,
        month = dec,
       volume = {391},
        pages = {481-506},
          doi = {10.1111/j.1365-2966.2008.13805.x},
archivePrefix = {arXiv},
       eprint = {0808.1227},
 primaryClass = {astro-ph},
       adsurl = {https://ui.adsabs.harvard.edu/abs/2008MNRAS.391..481S}
}

@ARTICLE{lamastra2010,
       author = {{Lamastra}, A. and {Menci}, N. and {Maiolino}, R. and {Fiore}, F. and {Merloni}, A.},
        title = "{The Building Up of the Black Hole--Stellar Mass Relation}",
      journal = {\mnras},
         year = 2010,
        month = jun,
       volume = {405},
        pages = {29-40},
          doi = {10.1111/j.1365-2966.2010.16439.x},
       adsurl = {https://ui.adsabs.harvard.edu/abs/2010MNRAS.405...29L}
}

@ARTICLE{barausse2012,
       author = {{Barausse}, E.},
        title = "{The Evolution of Massive Black Holes and Their Spins in Their Galactic Hosts}",
      journal = {\mnras},
         year = 2012,
        month = jul,
       volume = {423},
        pages = {2533-2557},
          doi = {10.1111/j.1365-2966.2012.21057.x},
archivePrefix = {arXiv},
       eprint = {1201.5888},
 primaryClass = {astro-ph.CO},
       adsurl = {https://ui.adsabs.harvard.edu/abs/2012MNRAS.423.2533B}
}

@ARTICLE{contini2026BH,
       author = {{Contini}, Emanuele and {Jang}, J.~K. and {Rhee}, Jinsu and {Seo}, Changjo and {Yi}, Sukyoung K.},
        title = "{Black Hole Occupation Fraction: Dependence on Black Hole Mass Threshold, Environment, Resolution and Redshift}",
      journal = {arXiv e-prints},
         year = 2026,
        month = jun,
          eid = {arXiv:2606.22758},
        pages = {arXiv:2606.22758},
          doi = {10.48550/arXiv.2606.22758},
archivePrefix = {arXiv},
       eprint = {2606.22758},
 primaryClass = {astro-ph.GA},
       adsurl = {https://ui.adsabs.harvard.edu/abs/2026arXiv260622758C}
}

@ARTICLE{bonoli2025,
       author = {{Bonoli}, Silvia and {Izquierdo-Villalba}, David and {Spinoso}, Daniele and {Colpi}, Monica and {Sesana}, Alberto and {Polkas}, Markos and {Springel}, Volker},
        title = "{Constraints on the early growth of massive black holes from PTA and JWST with L-GalaxiesBH}",
      journal = {arXiv e-prints},
         year = 2025,
        month = sep,
          eid = {arXiv:2509.12325},
        pages = {arXiv:2509.12325},
          doi = {10.48550/arXiv.2509.12325},
archivePrefix = {arXiv},
       eprint = {2509.12325},
 primaryClass = {astro-ph.GA},
       adsurl = {https://ui.adsabs.harvard.edu/abs/2025arXiv250912325B}
}

@ARTICLE{spinoso2023,
       author = {{Spinoso}, D. and {Bonoli}, S. and {Valiante}, R. and {Schneider}, R. and {Izquierdo-Villalba}, D.},
        title = "{Multiflavour SMBH seeding and evolution in cosmological environments}",
      journal = {\mnras},
         year = 2023,
        month = jan,
       volume = {518},
       number = {3},
        pages = {4672-4692},
          doi = {10.1093/mnras/stac3169},
archivePrefix = {arXiv},
       eprint = {2203.13846},
 primaryClass = {astro-ph.GA},
       adsurl = {https://ui.adsabs.harvard.edu/abs/2023MNRAS.518.4672S}
}

@ARTICLE{trinca2022,
       author = {{Trinca}, Alessandro and {Schneider}, Raffaella and {Valiante}, Rosa and {Graziani}, Luca and {Zappacosta}, Luca and {Shankar}, Francesco},
        title = "{The low-end of the black hole mass function at cosmic dawn}",
      journal = {\mnras},
         year = 2022,
        month = mar,
       volume = {511},
       number = {1},
        pages = {616-640},
          doi = {10.1093/mnras/stac062},
archivePrefix = {arXiv},
       eprint = {2201.02630},
 primaryClass = {astro-ph.GA},
       adsurl = {https://ui.adsabs.harvard.edu/abs/2022MNRAS.511..616T}
}

@ARTICLE{dayal2018,
       author = {{Dayal}, Pratika and {Ferrara}, Andrea},
        title = "{Early galaxy formation and its large-scale effects}",
      journal = {\physrep},
         year = 2018,
        month = dec,
       volume = {780},
        pages = {1-64},
          doi = {10.1016/j.physrep.2018.10.002},
archivePrefix = {arXiv},
       eprint = {1809.09136},
 primaryClass = {astro-ph.GA},
       adsurl = {https://ui.adsabs.harvard.edu/abs/2018PhR...780....1D}
}

@ARTICLE{volonteri-natarajan2009,
       author = {{Volonteri}, Marta and {Natarajan}, Priyamvada},
        title = "{Journey to the M$_{BH}$-{\ensuremath{\sigma}} relation: the fate of low-mass black holes in the Universe}",
      journal = {\mnras},
         year = 2009,
        month = dec,
       volume = {400},
       number = {4},
        pages = {1911-1918},
          doi = {10.1111/j.1365-2966.2009.15577.x},
archivePrefix = {arXiv},
       eprint = {0903.2262},
 primaryClass = {astro-ph.CO},
       adsurl = {https://ui.adsabs.harvard.edu/abs/2009MNRAS.400.1911V}
}

@ARTICLE{reines2015,
       author = {{Reines}, Amy E. and {Volonteri}, Marta},
        title = "{Relations between Central Black Hole Mass and Total Galaxy Stellar Mass in the Local Universe}",
      journal = {\apj},
         year = 2015,
        month = nov,
       volume = {813},
       number = {2},
          eid = {82},
        pages = {82},
          doi = {10.1088/0004-637X/813/2/82},
archivePrefix = {arXiv},
       eprint = {1508.06274},
 primaryClass = {astro-ph.GA},
       adsurl = {https://ui.adsabs.harvard.edu/abs/2015ApJ...813...82R}
}

@ARTICLE{wang2021,
       author = {{Wang}, Feige and {Yang}, Jinyi and {Fan}, Xiaohui and {Hennawi}, Joseph F. and {Barth}, Aaron J. and {Banados}, Eduardo and {Bian}, Fuyan and {Boutsia}, Konstantina and {Connor}, Thomas and {Davies}, Frederick B. and {Decarli}, Roberto and {Eilers}, Anna-Christina and {Farina}, Emanuele Paolo and {Green}, Richard and {Jiang}, Linhua and {Li}, Jiang-Tao and {Mazzucchelli}, Chiara and {Nanni}, Riccardo and {Schindler}, Jan-Torge and {Venemans}, Bram and {Walter}, Fabian and {Wu}, Xue-Bing and {Yue}, Minghao},
        title = "{A Luminous Quasar at Redshift 7.642}",
      journal = {\apjl},
         year = 2021,
        month = jan,
       volume = {907},
       number = {1},
          eid = {L1},
        pages = {L1},
          doi = {10.3847/2041-8213/abd8c6},
archivePrefix = {arXiv},
       eprint = {2101.03179},
 primaryClass = {astro-ph.GA},
       adsurl = {https://ui.adsabs.harvard.edu/abs/2021ApJ...907L...1W}
}

@ARTICLE{banados2018,
       author = {{Ba{\~n}ados}, Eduardo and {Venemans}, Bram P. and {Mazzucchelli}, Chiara and {Farina}, Emanuele P. and {Walter}, Fabian and {Wang}, Feige and {Decarli}, Roberto and {Stern}, Daniel and {Fan}, Xiaohui and {Davies}, Frederick B. and {Hennawi}, Joseph F. and {Simcoe}, Robert A. and {Turner}, Monica L. and {Rix}, Hans-Walter and {Yang}, Jinyi and {Kelson}, Daniel D. and {Rudie}, Gwen C. and {Winters}, Jan Martin},
        title = "{An 800-million-solar-mass black hole in a significantly neutral Universe at a redshift of 7.5}",
      journal = {\nat},
         year = 2018,
        month = jan,
       volume = {553},
       number = {7689},
        pages = {473-476},
          doi = {10.1038/nature25180},
archivePrefix = {arXiv},
       eprint = {1712.01860},
 primaryClass = {astro-ph.GA},
       adsurl = {https://ui.adsabs.harvard.edu/abs/2018Natur.553..473B}
}

@ARTICLE{natarajan2024,
       author = {{Natarajan}, Priyamvada and {Pacucci}, Fabio and {Ricarte}, Angelo and {Bogd{\'a}n}, {\'A}kos and {Goulding}, Andy D. and {Cappelluti}, Nico},
        title = "{First Detection of an Overmassive Black Hole Galaxy UHZ1: Evidence for Heavy Black Hole Seed Formation from Direct Collapse}",
      journal = {\apjl},
         year = 2024,
        month = jan,
       volume = {960},
       number = {1},
          eid = {L1},
        pages = {L1},
          doi = {10.3847/2041-8213/ad0e76},
archivePrefix = {arXiv},
       eprint = {2308.02654},
 primaryClass = {astro-ph.HE},
       adsurl = {https://ui.adsabs.harvard.edu/abs/2024ApJ...960L...1N}
}

@ARTICLE{maiolino2024,
       author = {{Maiolino}, Roberto and {Scholtz}, Jan and {Witstok}, Joris and {Carniani}, Stefano and {D'Eugenio}, Francesco and {de Graaff}, Anna and {{\"U}bler}, Hannah and {Tacchella}, Sandro and {Curtis-Lake}, Emma and {Arribas}, Santiago and {Bunker}, Andrew and {Charlot}, St{\'e}phane and {Chevallard}, Jacopo and {Curti}, Mirko and {Looser}, Tobias J. and {Maseda}, Michael V. and {Rawle}, Timothy D. and {Rodr{\'\i}guez del Pino}, Bruno and {Willott}, Chris J. and {Egami}, Eiichi and {Eisenstein}, Daniel J. and {Hainline}, Kevin N. and {Robertson}, Brant and {Williams}, Christina C. and {Willmer}, Christopher N.~A. and {Baker}, William M. and {Boyett}, Kristan and {DeCoursey}, Christa and {Fabian}, Andrew C. and {Helton}, Jakob M. and {Ji}, Zhiyuan and {Jones}, Gareth C. and {Kumari}, Nimisha and {Laporte}, Nicolas and {Nelson}, Erica J. and {Perna}, Michele and {Sandles}, Lester and {Shivaei}, Irene and {Sun}, Fengwu},
        title = "{A small and vigorous black hole in the early Universe}",
      journal = {\nat},
         year = 2024,
        month = mar,
       volume = {627},
       number = {8002},
        pages = {59-63},
          doi = {10.1038/s41586-024-07052-5},
archivePrefix = {arXiv},
       eprint = {2305.12492},
 primaryClass = {astro-ph.GA},
       adsurl = {https://ui.adsabs.harvard.edu/abs/2024Natur.627...59M}
}

@ARTICLE{larson2023,
       author = {{Larson}, Rebecca L. and {Finkelstein}, Steven L. and {Kocevski}, Dale D. and {Hutchison}, Taylor A. and {Trump}, Jonathan R. and {Arrabal Haro}, Pablo and {Bromm}, Volker and {Cleri}, Nikko J. and {Dickinson}, Mark and {Fujimoto}, Seiji and {Kartaltepe}, Jeyhan S. and {Koekemoer}, Anton M. and {Papovich}, Casey and {Pirzkal}, Nor and {Tacchella}, Sandro and {Zavala}, Jorge A. and {Bagley}, Micaela and {Behroozi}, Peter and {Champagne}, Jaclyn B. and {Cole}, Justin W. and {Jung}, Intae and {Morales}, Alexa M. and {Yang}, Guang and {Zhang}, Haowen and {Zitrin}, Adi and {Amor{\'\i}n}, Ricardo O. and {Burgarella}, Denis and {Casey}, Caitlin M. and {Ch{\'a}vez Ortiz}, {\'O}scar A. and {Cox}, Isabella G. and {Chworowsky}, Katherine and {Fontana}, Adriano and {Gawiser}, Eric and {Grazian}, Andrea and {Grogin}, Norman A. and {Harish}, Santosh and {Hathi}, Nimish P. and {Hirschmann}, Michaela and {Holwerda}, Benne W. and {Juneau}, St{\'e}phanie and {Leung}, Gene C.~K. and {Lucas}, Ray A. and {McGrath}, Elizabeth J. and {P{\'e}rez-Gonz{\'a}lez}, Pablo G. and {Rigby}, Jane R. and {Seill{\'e}}, Lise-Marie and {Simons}, Raymond C. and {de La Vega}, Alexander and {Weiner}, Benjamin J. and {Wilkins}, Stephen M. and {Yung}, L.~Y. Aaron and {Ceers Team}},
        title = "{A CEERS Discovery of an Accreting Supermassive Black Hole 570 Myr after the Big Bang: Identifying a Progenitor of Massive z > 6 Quasars}",
      journal = {\apjl},
         year = 2023,
        month = aug,
       volume = {953},
       number = {2},
          eid = {L29},
        pages = {L29},
          doi = {10.3847/2041-8213/ace619},
archivePrefix = {arXiv},
       eprint = {2303.08918},
 primaryClass = {astro-ph.GA},
       adsurl = {https://ui.adsabs.harvard.edu/abs/2023ApJ...953L..29L}
}

@ARTICLE{haidar2022,
       author = {{Haidar}, Houda and {Habouzit}, M{\'e}lanie and {Volonteri}, Marta and {Mezcua}, Mar and {Greene}, Jenny and {Neumayer}, Nadine and {Angl{\'e}s-Alc{\'a}zar}, Daniel and {Martin-Navarro}, Ignacio and {Hoyer}, Nils and {Dubois}, Yohan and {Dav{\'e}}, Romeel},
        title = "{The black hole population in low-mass galaxies in large-scale cosmological simulations}",
      journal = {\mnras},
         year = 2022,
        month = aug,
       volume = {514},
       number = {4},
        pages = {4912-4931},
          doi = {10.1093/mnras/stac1659},
archivePrefix = {arXiv},
       eprint = {2201.09888},
 primaryClass = {astro-ph.GA},
       adsurl = {https://ui.adsabs.harvard.edu/abs/2022MNRAS.514.4912H}
}

@ARTICLE{dubois2012,
       author = {{Dubois}, Yohan and {Devriendt}, Julien and {Slyz}, Adrianne and {Teyssier}, Romain},
        title = "{Self-regulated growth of supermassive black holes by a dual jet-heating active galactic nucleus feedback mechanism: methods, tests and implications for cosmological simulations}",
      journal = {\mnras},
         year = 2012,
        month = mar,
       volume = {420},
       number = {3},
        pages = {2662-2683},
          doi = {10.1111/j.1365-2966.2011.20236.x},
archivePrefix = {arXiv},
       eprint = {1108.0110},
 primaryClass = {astro-ph.CO},
       adsurl = {https://ui.adsabs.harvard.edu/abs/2012MNRAS.420.2662D}
}

@ARTICLE{dubois2014,
       author = {{Dubois}, Y. and {Pichon}, C. and {Welker}, C. and {Le Borgne}, D. and {Devriendt}, J. and {Laigle}, C. and {Codis}, S. and {Pogosyan}, D. and {Arnouts}, S. and {Benabed}, K. and {Bertin}, E. and {Blaizot}, J. and {Bouchet}, F. and {Cardoso}, J.-F. and {Colombi}, S. and {de Lapparent}, V. and {Desjacques}, V. and {Gavazzi}, R. and {Kassin}, S. and {Kimm}, T. and {McCracken}, H. and {Milliard}, B. and {Peirani}, S. and {Prunet}, S. and {Rouberol}, S. and {Silk}, J. and {Slyz}, A. and {Sousbie}, T. and {Teyssier}, R. and {Tresse}, L. and {Treyer}, M. and {Vibert}, D. and {Volonteri}, M.},
        title = "{Dancing in the dark: galactic properties trace spin swings along the cosmic web}",
      journal = {\mnras},
         year = 2014,
        month = oct,
       volume = {444},
       number = {2},
        pages = {1453-1468},
          doi = {10.1093/mnras/stu1227},
archivePrefix = {arXiv},
       eprint = {1402.1165},
 primaryClass = {astro-ph.CO},
       adsurl = {https://ui.adsabs.harvard.edu/abs/2014MNRAS.444.1453D}
}

@ARTICLE{han2025a,
       author = {{Han}, San and {Dubois}, Yohan and {Lee}, Jaehyun and {Kim}, Juhan and {Cadiou}, Corentin and {Yi}, Sukyoung K.},
        title = "{RAMSES-yOMP: Performance Optimizations for the Astrophysical Hydrodynamic Simulation Code RAMSES}",
      journal = {\apj},
         year = 2025,
        month = jan,
       volume = {978},
       number = {1},
          eid = {96},
        pages = {96},
          doi = {10.3847/1538-4357/ad98f4},
archivePrefix = {arXiv},
       eprint = {2411.14631},
 primaryClass = {astro-ph.IM},
       adsurl = {https://ui.adsabs.harvard.edu/abs/2025ApJ...978...96H}
}
\bibliographystyle{aasjournalv7}



\end{document}